\documentclass[preprint,12pt]{elsarticle}
\usepackage[utf8]{inputenc}
\usepackage[T1]{fontenc}

\usepackage[margin=2.25cm]{geometry}
\usepackage{amssymb}
\usepackage{hyperref}
\usepackage{amsmath}
\usepackage{braket}
\usepackage{soul}
\usepackage{xcolor}
\usepackage{enumitem}

\journal{Your Journal Name}
\begin{document}

\begin{frontmatter}
\title{Performance Analysis of Double Perovskite-Based Solar Cells Using SCAPS-1D Simulation: A brief review}

\author[inst1]{H. Laltlanmawii}
\author[inst1]{Lalrem Kima}
\author[inst2]{Mahabur Rahman}
\author[inst2]{Md. Ferdous Rahman}
\author[inst3,inst8]{Dilshod Nematov}
\author[inst4]{S. Bhattarai}
\author[inst7]{C. V. M. Chaturvedi}
\author[inst5]{Yazen M. Alawaideh}
\author[inst6]{A. Laref}

\affiliation[inst1]{organization={Advanced Functional Materials \& Simulation Lab (AFMSL), Department of Physics},Department and Organization
	addressline={Mizoram University}, 
	city={Aizawl},
	postcode={796004}, 
	country={India}}
\affiliation[inst2]{organization={Advanced Energy Materials and Solar Cell Research Laboratory}, Department and Organisation
	addressline={Department of Electrical and Electronic Engineering, Begum Rokeya University}, 
	city={Rangpur},
	postcode={ 5400}, 
	country={Bangladesh}}

\affiliation[inst3]{organization={Physics}, Department and Organization addressline={S.U. Umarov Physical-Technical Institute of the National Academy of Sciences of Tajikistan (NAST)}, 	city={Dushanbe}, postcode={734063}, country={Tajikistan}}
\affiliation[inst8]{organization={School of Optoelectronic Engineering \& CQUPT-BUL Innovation Institute}, Department and Organization addressline={Chongqing University of Posts and Telecommunications}, 	city={Chongqing}, postcode={400065}, country={China}}
\affiliation[inst4]{organization={Technology Innovation and Hub}, Department and Organization addressline={Indian Institute of Technology Guwahati}, 	city={, Guwahati, Assam}, postcode={792103}, country={India}
}

\affiliation[inst7]{organization={Department of ECE}, Department and Organization addressline={GVP College of Engineering (Autonomous)}, 	city={Madhurawada, Visakhapatnam, Andhra Pradesh}, postcode={530048}, country={India}
}

\affiliation[inst5]{organization={Department of Computer Engineering}, Department and Organization addressline={Biruni University}, 	city={Istanbul}, postcode={34010}, country={Turkey}}

 \affiliation[inst6]{organization={Department of Physics and Astronomy, College of Science}, addressline={King Saud University}, 
	city={Riyadh},
	postcode={11451}, 
	country={Saudi Arabia}}
\author[inst1]{D. P. Rai\texorpdfstring{\corref{cor1}}{}}
\cortext[cor1]{dibyaprakashrai@gmail.com}  

\begin{abstract}
Lead-free double perovskites are among the rapidly developing next-generation solar cell technologies, providing the required low toxicity, stability, as well as high optoelectronic potential. So far, experimentally prepared lead-free perovskite solar cell devices are reported to have low power conversion efficiency (PCE) for practical application as compared to the lead-based perovskites. However, the experimental preparation of lead-free perovskite supercells and optimize their PCE from the large pool of combinatorial compositions is challenging, which demands a large amount of time and money. In recent years, numerical simulations have emerged as a cost-effective approach that plays a crucial role in expediting scientific research, can bridge the gap between experiment and theory, and provide predictive information regarding the preparation of solar cells and their PCEs without undergoing real-time experiments. The tools, such as 1D numerical simulation software SCAPS--1D (Solar Cell Capacitance Simulator), are now needed to test newer architectures and determine what exactly is holding them back. An in-depth discussion of these systems, with great potential, is presented in this paper and simulated in the optical simulator SCAPS-1D. We subdivide the analysis into four main points: bandgap analysis of all absorbers, role of Electron Transport Layers (ETLs), role of absorber properties, and the role of Hole Transport Layers (HTLs). The long-term simulation results show that the best PV bandgap window for these absorbers is 1.5 to 1.8 eV. The simulated configurations clearly indicate that it is critical to obtain bulk defect densities below 10$^{15}$ cm$^{-3}$ and absorber thicknesses from 500 to 900 nm. The incorporation of high-mobility transport layers greatly reduces interfacial recombination, resulting in power conversion efficiencies over 32\%. In this review article, we try to highlight the comprehensive analysis of SCAPS-1D in terms of merits and demerits, focusing on the perovskite materials. So far in the field of solar cell research, SCAPS-1D has been extensively used and looks like a powerful software due to its user-friendliness and simulation of results in a few seconds. The speed and ease of simulation make SCAPS-1D a very popular tool; as a result, it enables rapid optimization of a large number of photovoltaic devices and their performances without undergoing any experimental work, which can save time and money. However, one serious drawback is that the SCAPS-1D simulator works only for 1D configurations. It is ineffective in incorporating atomistic interactions and 3D effects. Hence, the efficacy of the SCAPS-1D simulator solely relies on the accuracy of the input parameters (input data from DFT or experiment) that the user provides, failing which may give wrong results and large deviations from accuracy.  
\end{abstract}

\end{frontmatter}

\newpage																		
\tableofcontents 
\section[Introduction]{Introduction}
\begin{figure}[h!]
    \centering
    \includegraphics[width=0.80\linewidth]{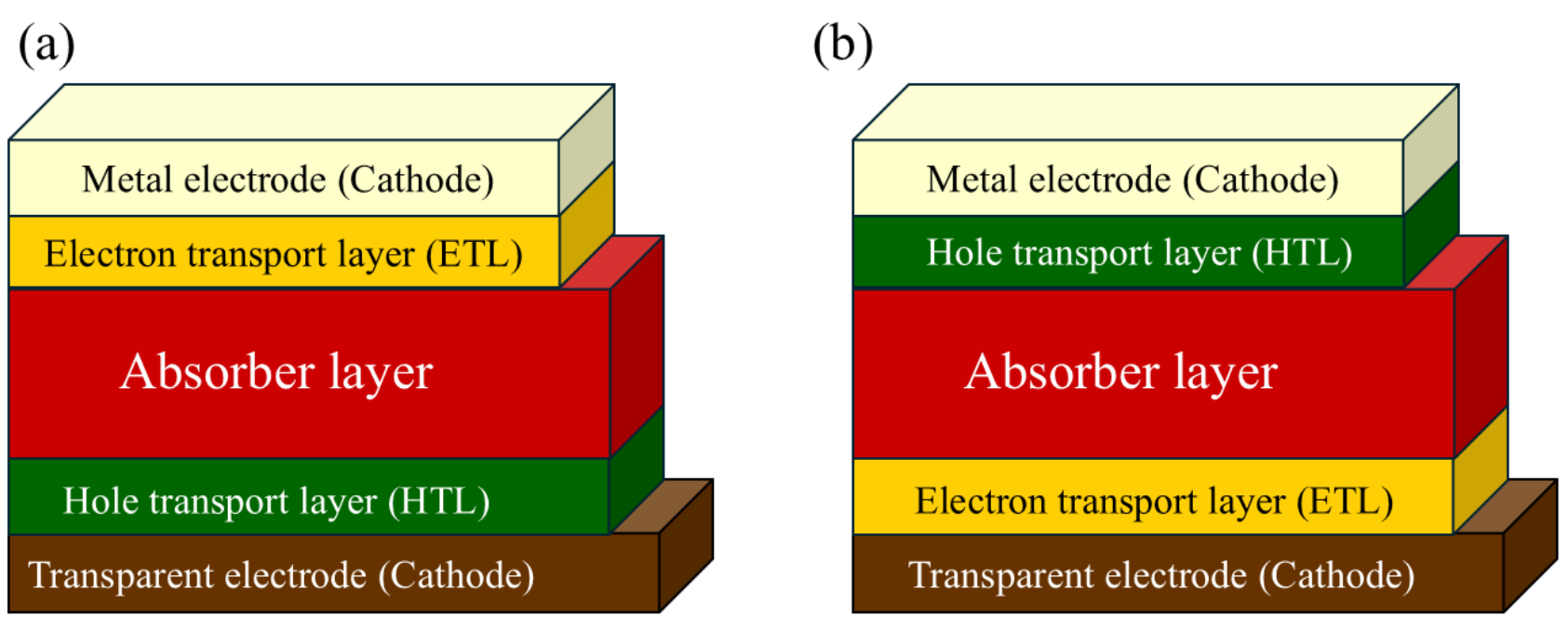}
    \caption{Sandwich model of solar cell}
    \label{fig.1}
\end{figure}

The current scientific research is directed towards the development of clean and sustainable energy-harvesting systems due to the rapid depletion of conventional energy resources (fossil fuels) and their significant contribution to damaging the ecosystem via environmental pollution by greenhouse gas emissions.\cite{FEIGIN2023}\cite{JAISWAL2022} The production of usable electrical energy by trapping sunlight using a solid-state semiconductor panel is the most promising and widely applicable solution to the global energy crisis. The sandwich-layered architecture of a solar cell comprises components like a hole-transport layer (HTL), electron-transport-layer (ETL), absorber layer and metal contacts that work on the charge transfer mechanism, giving rise to the photovoltaic effect [see Fig.\ref{fig.1}]. Solar cells promise clean, sustainable, and affordable energy. As a result, the global demand for solar cell panels/wafers is increasing year by year.  However, despite having extensive research in this field, it is suffering from several deficiencies such as problems in energy storage, low efficiency (30\%$<$) for real-time applications, occupying a large area, non-degradable solar panels, etc.\cite{Ebhota2023, Kung2020}. Researchers are committed to understanding the underlying physics to overcome the current limitations of the solar cell by adopting various strategies, including bandgap engineering, strain-stress effect, doping, modulating layer numbers, etc.

\par Solar cells are applicable in all sectors for power generation:
\begin{itemize}
    \item for large-scale power generation in factories, universities, government offices, automobiles, sports arenas, spaceships, aerospace, etc., via dense-grid technology.\cite{LIU2025, GHOSH2025}
    \item small power generation in houses, public buildings, schools, electronic devices, sensors, street lights,  Internet of Things (IoT), etc.\cite{BHAU2023,ELHAMMOUMI2022}
\end{itemize} 
\par
\begin{figure}[h!]
    \centering
    \includegraphics[width=0.80\linewidth]{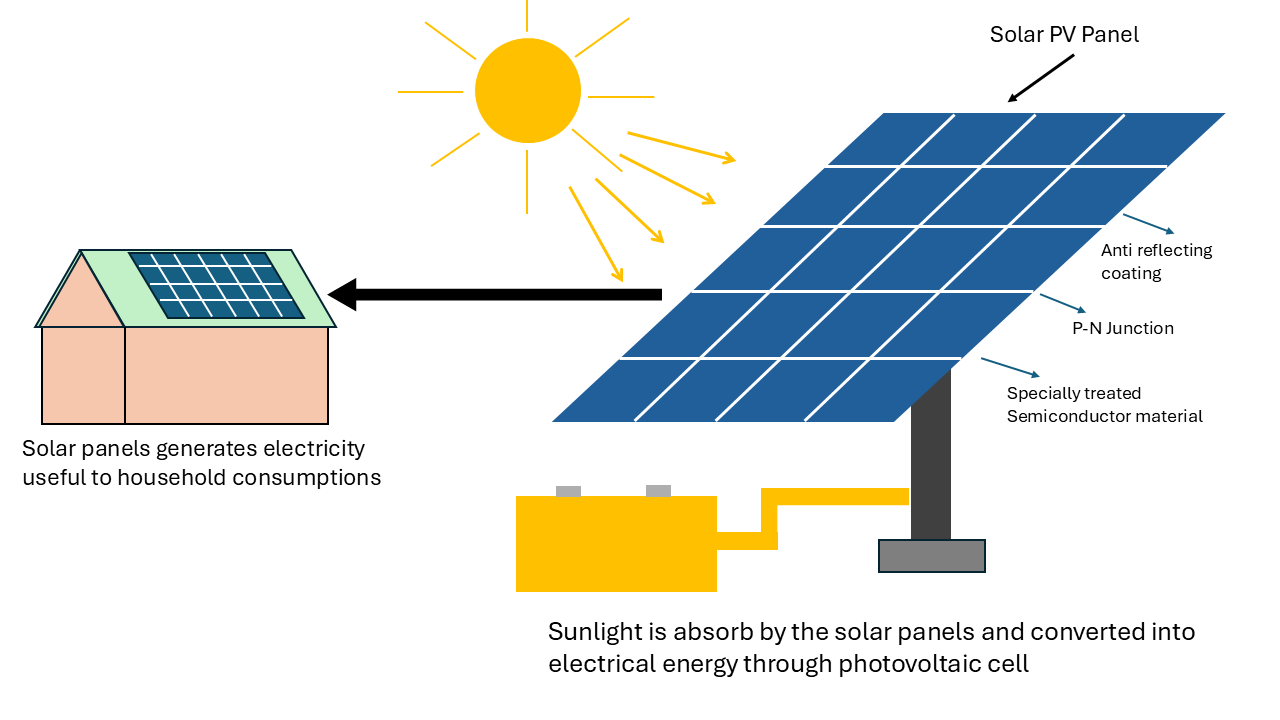}
    \caption{Solar panel in everyday use for generating electricity}
    \label{fig.2}
\end{figure}

 The solar cell devices are fabricated from different types of semiconductor materials such as silicon\cite{MEENA2024}, perovskite\cite{Hima2025,SANAP2024}, organic-semiconductor\cite{MATAKGANE2023}, sulvanite\cite{LALROLIANA2024}, dye-sensitized materials\cite{AGRAWAL2022}, delafossite\cite{Laltlanmawii2025}, kesterite\cite{Hani2026}, etc. Each one of them has unique characteristics (advantages/merits and disadvantages/drawbacks). It is crucial to understand how different solar cells work under different conditions, their efficiencies and the cost of production. A thorough analysis can help us identify the most efficient solar cell for a target-specific application. This work will mostly focus on the emerging alternative perovskite materials for their compatibility in solar cell applications, with the purpose of understanding their performance for the next-generation solar cell devices. A solar cell with high efficiency is desirable; however, it is necessary to consider affordability in terms of its cost. Another point to address is environmental friendliness, as many of the Pb-based solar cells have high efficiency, but disposing of them after expiry leads to soil pollution. By addressing all these crucial research points, the review aims to bridge the gap between materials selection, real-time laboratory fabrication of environment-friendly solar cells and their conditional applications. Fig.\ref{fig.2} shows the application of solar cell technology in our daily life. 

\section{Semiconductor Technology}
The study of semiconductor materials was initiated by Michael Faraday's discovery in 1833 and steadily progressed towards the invention of the first silicon transistor in 1954, signifying the dawn of contemporary silicon electronics and microelectronics. Before the development of band theory, advancements in this area were gradual and primarily driven by technological demands. The advent of band theory, alongside dedicated efforts to cultivate high-purity silicon and germanium and to comprehend their intrinsic properties, spurred a swift and ongoing expansion of semiconductor research that persists to this day.

\par A significant advancement occurred during the 1930s and 1940s with the advent of quantum band theory, which clearly explained semiconductor conductivity, band gaps, and the effects of doping. This theoretical groundwork paved the way for the creation of functional semiconductor devices. The field transformed in 1947 when Bardeen, Brattain, and Shockley introduced the first transistor at Bell Labs, signalling the dawn of modern electronics. The shift from germanium to silicon in the early 1950s, especially with the introduction of the first silicon transistor in 1954, initiated the age of silicon-based microelectronics. The invention of the integrated circuit between 1958 and 1959, along with subsequent progress in lithography and materials science, established semiconductors as the foundation of all current electronic and optoelectronic technologies.

\par In 1839, A.E. Becquerel was experimenting with electrolytes. He worked with silver chloride-coated platinum placed in nitric acid, and noticed something that when sunlight hit one of the electrodes, the voltage (emf) of the electrodes increased. This made the first record of the photovoltaic effect. Later on, in 1873, Willoughby Smith discovered the photoconductivity of selenium; he mistakenly believed selenium was a metal. Later, it took 14 years for Hertz to observe photoconductivity in actual metals, and selenium was only definitely identified as a semiconductor in 1907. In 1876, William Grylls Adams and Richard Evans Day discovered that selenium produced electricity when light was shone on it. In 1883, a significant advancement in practical solar technology was made when Charles Fritts created the first solid-state solar cell. This was accomplished by layering a thin film of selenium with a semitransparent gold layer. Despite its efficiency being under 1\%, this device was the first operational semiconductor solar cell, demonstrating the possibility of direct light-to-electricity conversion. The early 20th century saw further exploration into semiconductors, with research into rectification, p-n junctions, and the purification of silicon and germanium establishing the theoretical and technological basis for contemporary photovoltaics. These developments reached a peak in the 1940s and 1950s, when R. S. Ohl and Bell Laboratories' scientists developed high-purity silicon and stable p-n junctions that enabled efficient photovoltaic function. This progress led to the creation of the first practical silicon solar cell in 1954, heralding the onset of modern photovoltaic technology.

Here are a few advantages 
\begin{enumerate}[label=\roman*.]
    \item Crystalline Silicon (Mono- / Polycrystalline)- High efficiency and mature technology. Long lifetime and stability. Well-understood manufacturing processes.
\item Amorphous Silicon- Very high optical; it is low cost and can be deposited at low temperatures on various substrates; it is lightweight and flexible, making it good for building-integrated PV or curved surfaces.
\item Micromorph silicon- Has very good stability, even better than Amorphous silicon. It reduces material use because of thin films.
\item CIGS (Copper Indium Gallium Selenide)- Very high absorption coefficient. Good efficiency potential; lab and module efficiencies are strong.\cite{Asif2025}
\item Multi-Junction- ideal for high performance or space applications. It can exceed the Shockley-Queisser Limit.
\end{enumerate}
\par
\begin{figure}[h]
    \centering
    \includegraphics[width=1.0\linewidth]{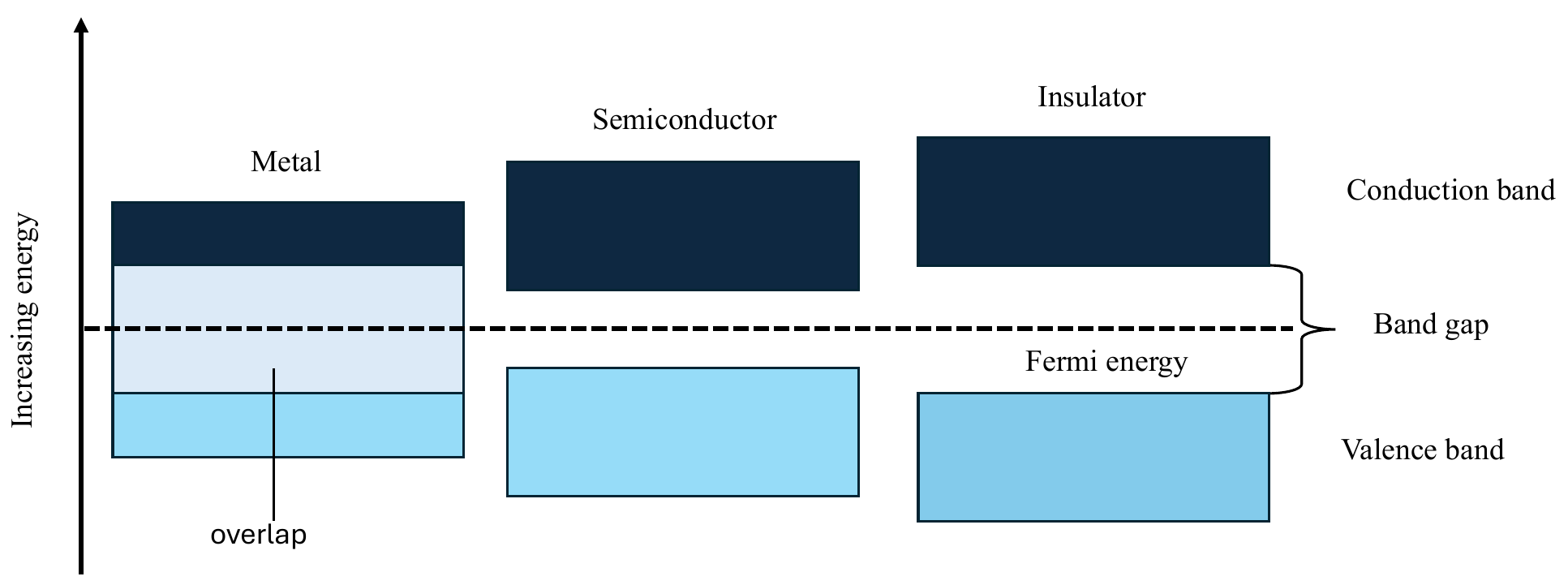}
    \caption{Band structure of different materials}
    \label{fig.3}
\end{figure}

 Semiconductor materials (Si, GaAs, CdTe, CIGS, etc.) play an important role in modern technology owing to their unique electronic properties. The band theory of solids can explain the behaviour of semiconductor materials. The band structures, band filling, and band gaps (Eg), which represent the energy required to jump from the valence band (VB) to the conduction band (CB)  (see Fig \ref{fig.3}), categorize a material into metal, semiconductor, or insulator.\cite{KUMAR2025} The Fermi energy is the highest energy state of a material at zero temperature, and for semiconductors and insulators it lies within the band gap, whereas for metals it is located within the CB. 

\section{Different types of solar cells} Fig\ref{fig.4} shows the different types of solar cell. They are classified into three generations as given below:
\subsection{First-generation solar cells} These are solar cells based on crystalline silicon wafers. These cells dominate the global photovoltaic market and achieve the highest efficiencies among commercial technologies due to exceptional performance and stability. Nearly 80\% of the world's PV production relies on single-crystalline silicon.\cite{Dambhare2021} The first generation is classified into two sub-groups: monocrystalline and polycrystalline. Monocrystalline Si currently delivers the highest efficiencies, with typical commercial modules reaching about 18–22\% and even higher values achieved in laboratory cells. Polycrystalline Si modules in commercial use typically show efficiencies of 15–20\%, lower than monocrystalline, but they offer cheaper manufacturing expenses.\cite{Rathore2021}

\subsection{Second-generation solar cells} Conventional silicon wafer-based solar cells are expensive because they require highly purified crystalline silicon, which involves a complex and costly manufacturing process. Thin-film technology reduces cost by depositing ultra-thin silicon layers around 1 $\mu$m thick, significantly lowering the amount of silicon required compared to wafer-based cells.\cite{BADAWY2015} Examples of thin-film solar cell (TFSC) materials include a-Si (amorphous silicon), CdTe (cadmium telluride), CIGS (copper indium gallium selenide), and CZTS (copper zinc tin sulfide). The a-Si generally has roughly 6–12\% efficiency, and CIGS laboratory cells achieve about 20–22\%, while commercial CdTe modules reach up to 21\%. CZTS currently show modest laboratory efficiencies in the 6–12\% range. Regardless of several advantages, the overall efficiency is only 7\%, impeding further commercialization.\cite{LEE2017}

\subsection{Third-generation solar cells} Third-generation solar cells aim to exceed the Shockley–Queisser efficiency limit and address the performance limits of crystalline silicon and thin-film technologies. They introduce advanced concepts and materials to achieve higher efficiency, lower cost, and better overall performance.\cite{DHILIPAN2022} This category includes several emerging high-efficiency solar cell types, namely, Dye-sensitized solar cells (DSSCs), Quantum-dot solar cells (QDSCs), Perovskite solar cells (PSCs), and organic solar cells. DSSCs achieve about 8--11\% efficiency in laboratory and small--area devices.  Certified practical demonstrations are currently around 10\%, with ideal MEG models predicting up to ~40\% for QDSCs. Organic solar cells currently show laboratory efficiencies of about 10-11\%, with module efficiencies somewhat lower and PSCs have progressed rapidly laboratory single-junction devices, now exceeding 20\% efficiency after only a few years of development.\cite{Soonmin2023, Wu2025} 

\begin{figure}[h]
    \centering
    \includegraphics[width=0.90\linewidth]{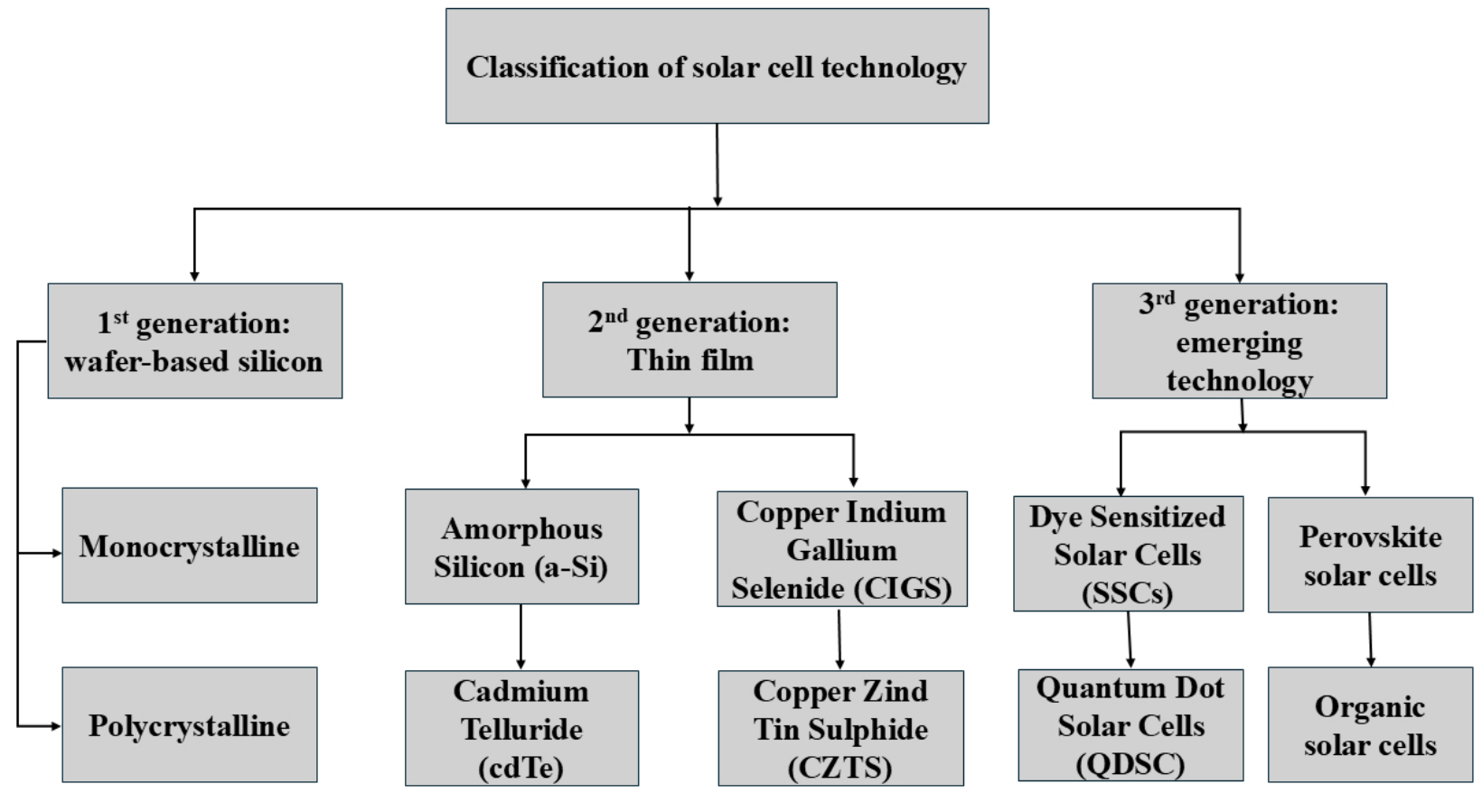}
    \caption{Different types of solar cells}
    \label{fig.4}
\end{figure}

\begin{figure}[h]
    \centering
    \includegraphics[width=0.80\linewidth]{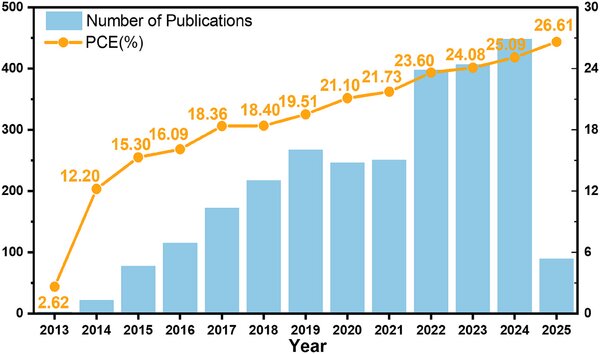}
    \caption{Year-wise number of publications and increase in their efficiencies of perovskite solar cells. Reprinted from \cite{Wu2025} with permission from Wiley.}
    \label{fig.5}
\end{figure}

\begin{figure*}[h]
    \centering
    \includegraphics[width=0.40\linewidth]{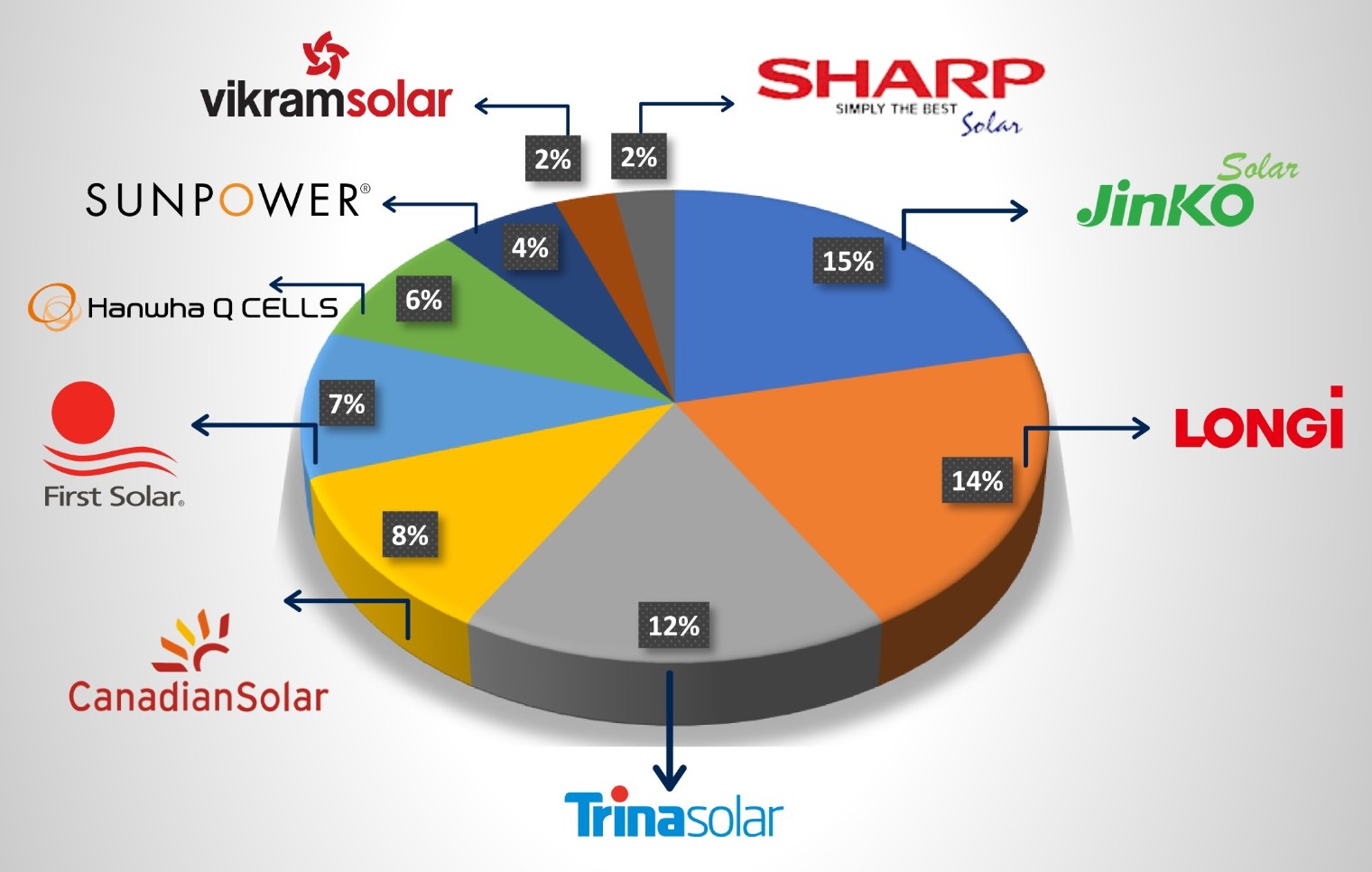}
        \includegraphics[width=0.50\linewidth]{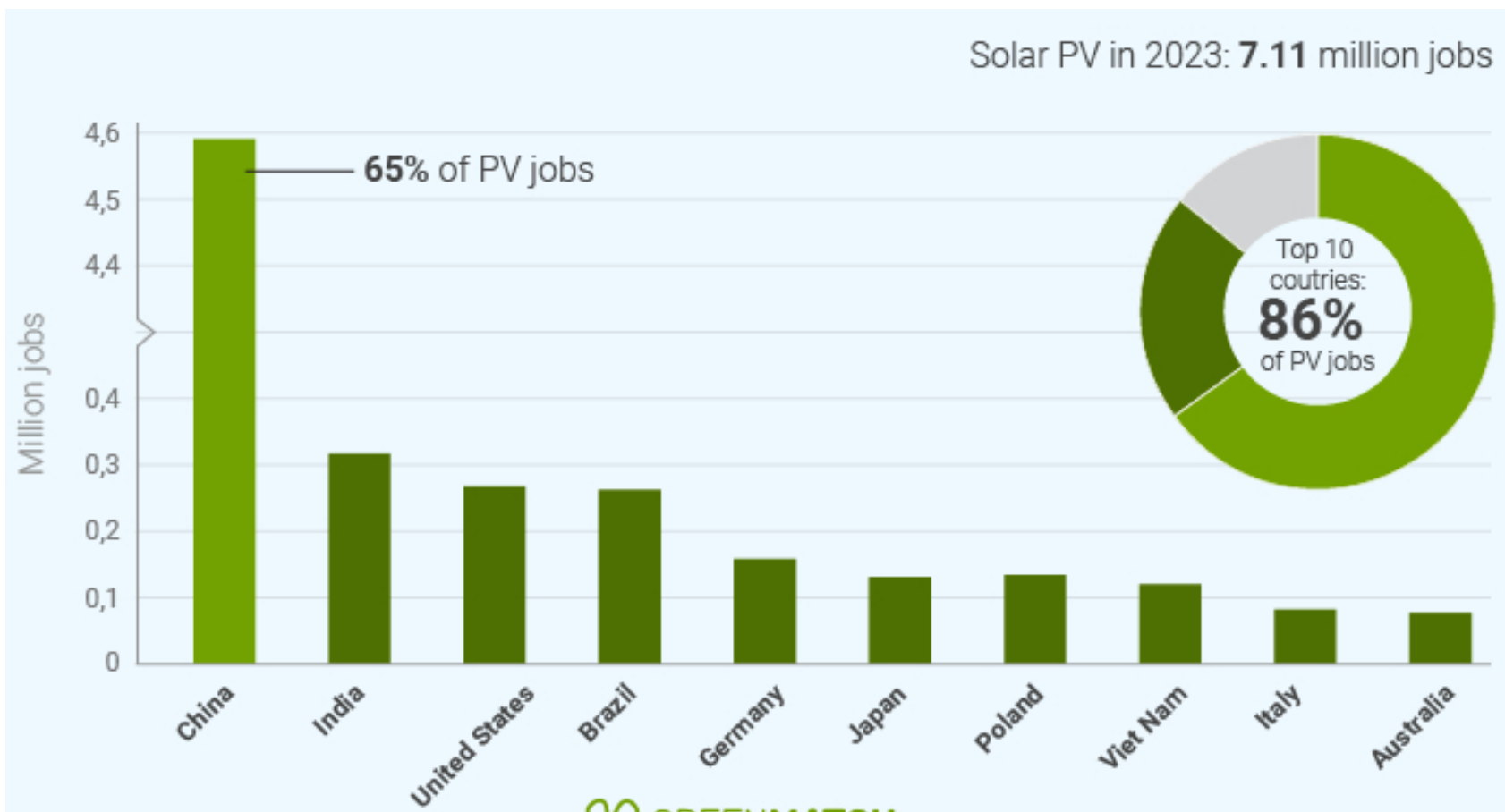}
    \caption{Companies involved in the production of perovskite solar cells and countries employed in the field of perovskite solar cells (Source: Forbes Magazine 23$^{rd}$ July 2019)}
    \label{fig.6}
\end{figure*}

\section{Perovskite Solar cell}
Perovskite materials have transformed the field of optoelectronics du to their tunability in band gaps, outstanding optoelectronic behaviour, and highly adaptable crystal structures. Modern PSCs have now achieved efficiencies of around 27\% (see Fig.\ref{fig.5}), positioning perovskites as strong contenders for next-generation photovoltaic technologies.\cite{Zhang2025} PSCs have rapidly gained attention, but their reliance on lead-based compounds raised serious concerns regarding toxicity and long-term environmental impact. They can reach efficiency levels close to the theoretical limit, but their commercial use is still limited, mainly because of poor stability and the presence of toxic materials such as lead. Perovskites can easily degrade when exposed to moisture, heat, or long periods of light under normal conditions. Secondly, lead-based perovskites have shown a tendency to release toxic \(PbI_2\) as a degradation product. However, replacing lead with a safer alternative remains one of the most important goals.\cite{Kung2020} Nevertheless, attempts like substituting \(Pb^{2+}\) with tin (\(Sn^{2+}\)) or germanium (\(Ge^{2+}\)) were unsuccessful due to rapid oxidation and poor stability.\cite{Noel2014, ROKNUZZAMAN2018} Later on the idea of substituting \(Pb^{2+}\) ions with a pair of non-toxic metals--one monovalent (+1) and one trivalent (+3) was produced, for example (\(Ag^+ + Bi^{3+}\)), (\(Na^+ + Sb^3+\)), and (\(Cu^+ + Bi^{3+}\)).\cite{AHN2020} Then the double perovskites, with the general formula \(A_2M(I)^+M(III)^{3+}X_2\), which is a lead-free material, were introduced.\cite{Liao2020} These materials offer better chemical stability, improved environmental safety, optoelectronic tunability, and, in some cases, can be synthesized with low cost and reduced complexity.  

Double perovskites are attracting significant interest as a promising category of lead-free materials because they effectively address the two primary drawbacks of traditional lead-based perovskite solar cells: toxicity and instability. By substituting divalent \(Pb^{2+}\) with a combination of heterovalent metal ions (usually \(M(I)^+\) and \(M(III)^{3+}\)), the use of hazardous lead is eliminated.\cite{Tiang2017} This change prevents the release of toxic \(PbI_2\) degradation products and avoids environmental contamination related to the dissolution of \(Pb^{2+}\) in water.\cite{Babayigit2016} This essential change in composition significantly enhances the environmental safety and sustainability of photovoltaic technology. Besides their non-toxic nature, halide double perovskites (HDPs), such as \(Cs_2AgBiBr_6\), demonstrate superior intrinsic chemical and thermal stability when compared to organic–inorganic lead perovskites, which are prone to degradation from moisture, heat, and extended exposure to light.\cite{Volonakis2016} Their robust inorganic framework, absence of easily oxidized species (such as \(Sn^{2+}\) or \(Ge^{2+}\)), and stronger metal–halide bonding contribute to significantly improved durability under operational conditions. Moreover, double perovskites retain the general perovskite crystal architecture, allowing flexibility in B-site cation substitution and enabling the exploration of numerous combinations (e.g., Ag–Bi, Ag–In, In–Sb) that may yield suitable band gaps and enhanced optoelectronic properties.\cite{McClure2016} This structural tunability opens pathways toward safer and potentially more stable photovoltaic materials, marking double perovskites as an appealing platform for next-generation lead-free solar cells.\cite{Chu2019} Several countries are investing huge budget in solar cell manufacturing sectors and some of the companies already involved in the production of commercial solar-cells are shown in Fig.\ref{fig.6} 

Early research on double perovskites focused mainly on oxide-based materials because they showed attractive properties such as paramagnetism, ferromagnetism, and magnetoresistance. Research has also shifted from oxide-based double perovskites, known for magnetic properties like ferromagnetism and magnetoresistance, to halide-based double perovskites, which are more suitable for PV applications. Double perovskites offer greater environmental safety, enhanced structural stability, and more flexibility in choosing metal cations. Although they still face challenges such as wide and often indirect band gaps, they provide a promising pathway toward stable, lead-free, next-generation perovskite solar cells.\cite{Kung2020}

Chalcogenide double perovskites have been suggested as a lead-free and potentially more stable option compared to HDPs like \(Cs_2AgBiBr_6\). However, they exhibit indirect band gaps, weak absorption, and low carrier mobility. Consequently, most chalcogenide double perovskites continue to face challenges in acting as efficient solar absorbers.\cite{Sun2019}. Also, lead-free (Pb-free) HDPs have gained popularity and promise due to their non-toxic nature, increased stability, and the potential for versatile chemical design. Nevertheless, according to current theoretical and experimental findings, which encompass structural, electronic, and optical analyses, the majority of Pb-free double perovskites encounter inherent limitations that hinder them from reaching efficiencies as high as those of the latest lead-based perovskites.\cite {Longo2020}.

\subsection{Development of Perovskite Solar Cell} In 1839, Gustav Rose first discovered calcium titanate (CaTiO$_3$) minerals in Russia, and later coined the term 'perovskite' for the structure after a Russian mineralogist Lev A. Perovski (1792-1856).\cite{SANGA2025} Throughout the years, other inorganic metal oxides ( BaTiO$_3$, SrTiO$_3$, PbTiO$_3$, etc.) having the same perovskite structures emerged, attracting much interest in the scientific community owing to their ferroelectric and piezoelectric applicability. Although a successful synthesis was carried out as early as 1893, the photovoltaic breakthrough for perovskite materials did not come to light until the early 21st century.\cite{Green2017} Research on the photovoltaic capabilities of perovskite began in 2005 at Miyasaka's laboratory at Toin University in Japan. In 2009, Kojima and colleagues reported a PCE of 3.81\% when utilizing methylammonium lead iodide (CH$_3$NH$_3$PbI$_3$) in photoelectrochemical cells, marking the beginning of perovskite solar cells (PSCs) and has since become the forefront of photovoltaic technology.\cite{Kojima2009} In 2012, a breakthrough happened when Gratzel and Park replaced the liquid electrolyte with a solid hole conductor (Spiro-OMeTAD), creating the first solid-state PSC\cite{Calio2016}. The record PCE for PSCs have exceeded 25\% for single-junction cells, while tandem cells have climbed up to 30\%.  

\subsection{Working Mechanism of Perovskite Solar Cell} 
\begin{figure*}[h!]
    \centering
    \includegraphics[width=0.45\linewidth]{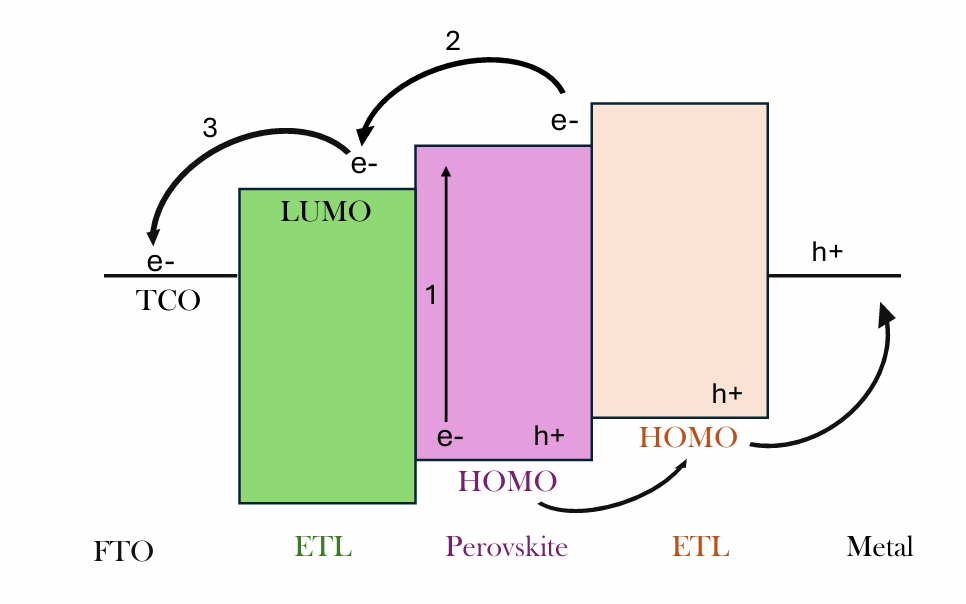}
        \includegraphics[width=0.45\linewidth]{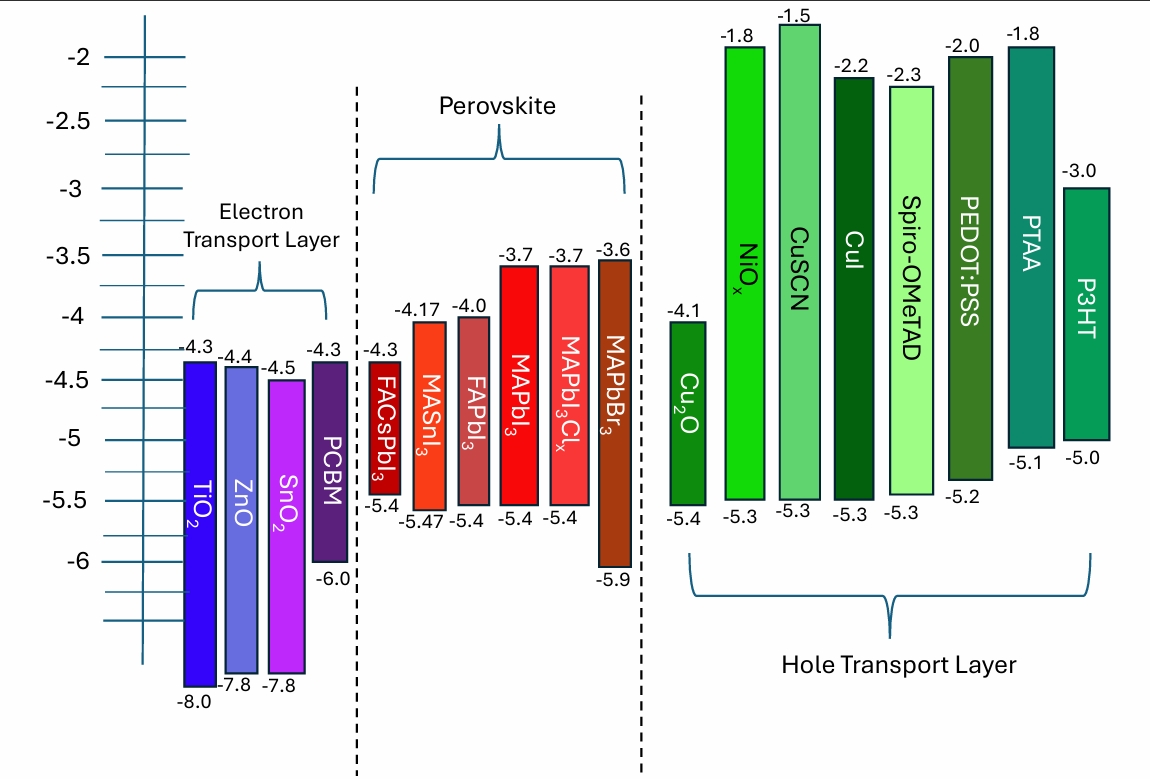}
    \caption{Working mechanism of perovskite solar cells}
    \label{fig.7}
\end{figure*}

The working mechanism of perovskite solar cells has been the subject of intensive research in recent years, and most studies emphasize that coordinated charge transfer through the absorber layer, ETL, and HTL plays a key role in achieving high efficiency. The energy diagram shown in Fig.\ref{fig.7} systematizes current understanding of charge transport in PSCs, illustrating the traditionally distinguished stages: generation, separation, and selective transport of electrons and holes. As noted in a number of recent works, \cite{Chouhan2020, Noman2024} perovskites have high absorption capacity and a narrow band gap ($\sim$1.5 eV), which ensures high efficiency of light conversion into charge carriers. The strength of this architecture is confirmed by the fact that, due to the low exciton binding energy ($\sim$2 meV), free carriers are formed almost immediately after photon absorption,\cite{Chang2022} which reduces energy losses and increases transport efficiency.

\par Researchers emphasize that spatial architecture and energy alignment of layers, similar to that shown in Fig.\ref{fig.7}, are determining factors in PSC efficiency. According to the results of \cite{Noman2024, Mohammadi2024}, the high diffusion length of carriers (up to hundreds of nanometers) contributes to the fact that electrons and holes can be effectively separated and delivered to the transport layers without significant recombination. Analysis of the literature shows that the transport characteristics of ETL and HTL play a key role in carrier dynamics: for example, Mujahid \textit{et al.} state that ETL materials such as TiO$_2$ and SnO$_2$ contribute differently to electron transport.\cite {Mujahid2025} TiO$_2$, despite its prevalence, exhibits relatively low mobility (0.1-4 cm$^2$/V s), while the use of SnO$_2$ significantly improves carrier extraction due to its higher mobility ($\sim$100 cm$^2$/V s).\cite{Mohammadi2024, KUMAR2022} These observations are consistent with the direction of electron transport shown by arrows (2) in Fig.\ref{fig.7}, where the energy barrier between perovskite and ETL is minimized.

\par On the other hand, research in the field of hole transport materials indicates the importance of matching the HOMO HTL and perovskite levels for efficient hole transport. For example, Mujahid \textit{et al.}\cite{Mujahid2025} and Wu \textit{et al.}\cite{Wu2021} show that Spiro-OMeTAD and CuSCN provide a higher level of selectivity for holes, while recent works, such as the study by Raj \textit{et al}.\cite{Raj2025}, demonstrate the advantages of Cu$_2$O, which allows for optimization of energy alignment and reduction of recombination probabilities. The simulation conducted in \cite{Raj2025} confirms that a small HOMO level shift between HTL and perovskite has a significant effect on V$_{oc}$ parameters, which is fully consistent with the hole transport trajectories represented by arrows (3) in Fig.\ref{fig.7}. Thus, analysis of recent publications confirms the fundamental role of HTL in selective hole drain and support for efficient device operation.
\par Current research confirms that one of the main factors limiting PSC efficiency is carrier recombination at interfaces.\cite{Chang2022, Mohammadi2024, Mujahid2025, KUMAR2022, Wu2021}. According to \cite{Chang2022} and \cite{Mohammadi2024}, the perovskite/ETL and perovskite/HTL interfaces are regions with increased trap density, which contributes to non-radiative recombination described by the Shockley-Reed-Hall model:
\begin{equation}
    R_{SRH}=\frac{np-n_1^2}{\tau_p(n+n_1)+\tau_n(p+p_1)}
\end{equation}
Works by \cite{Chang2022, Wu2021} confirm that the use of passivation additives and buffer layers can significantly reduce defect density and, accordingly, recombination rate. For example, the introduction of a thin layer of BiI$_3$ at the perovskite/HTL interface was proposed in \cite{Chang2022} and demonstrated a significant reduction in losses and an increase in efficiency, while Pt doping of the TiO$_2$ surface \cite{Mohammadi2024, KUMAR2022} increased the efficiency from 12.9\% to 14.4\%. These results show that reducing recombination at interfaces is one of the most effective ways to increase PSC efficiency and is directly related to the charge transfer mechanism shown in Fig.\ref{fig.7}, since interface traps hinder the selective movement of carriers, disrupting the physical processes represented by the arrows in the diagram.
\par A significant contribution to understanding the influence of contacts on the transport mechanism was made in \cite{Wu2021}, where it was shown that matching the operation of the TCO contact output with ETL and the back metal contact with HTL reduces barrier effects and increases transport efficiency. Studies conducted using Ni as the back contact have shown the advantages of this metal in blocking the reverse electron current, which is consistent with the direction of hole transport in Fig.\ref{fig.7}.
\par Existing data confirm that high efficiency is achieved through direct carrier generation upon photon absorption, energy alignment between perovskite, ETL, and HTL, selective electron and hole transport, and interface engineering aimed at reducing recombination. An analysis of recent literature shows that it is the combination of these factors that gives PSCs their high performance and makes them a promising platform for further improvements in efficiency and stability. 

\subsection{Synthesis of perovskite and Solar cell Device fabrication method}
The experimental techniques used in perovskite solar cell (PSC) research mainly include co-precipitation, hydrothermal, solid-state reaction, and sol-gel method.\cite{Afre2024} The chosen form of synthesis determine the specific properties of the particle, such as shape, size, and morphology. The co-precipitation method is simple, low-cost, and water-based, operating at a low reaction temperature ($50-80^\circ\text{C}$) while producing high yields and large particle. However, controlling size and morphology is reported to be challenging.\cite{FU2017} Similarly, the sol-gel method offers excellent chemical homogeneity, precise stoichiometric control, and high-purity products at lower temperature, although secondary impurity phases are occasionally produces which hinders the application of the material.\cite{BRUNCKOVA2016} The hydrothermal route requires sealed reactors at high temperatures and pressures (typically $160-250^\circ\text{C}$) to produce highly crystalline powders without calcination, while allowing the particle size and morphology to be altered by controlling the reactions.\cite{Esposito2019} In contrast, the solid-state reaction method requires high temperature calcination (over $1000^\circ\text{C}$), making it cost-effective and desirable but generally produce large particles with broad size distribution.\cite{OHARA2008} 

\section{Structural Properties of Perovskites}
\begin{figure}[h]
    \centering
    \includegraphics[width=0.95\linewidth]{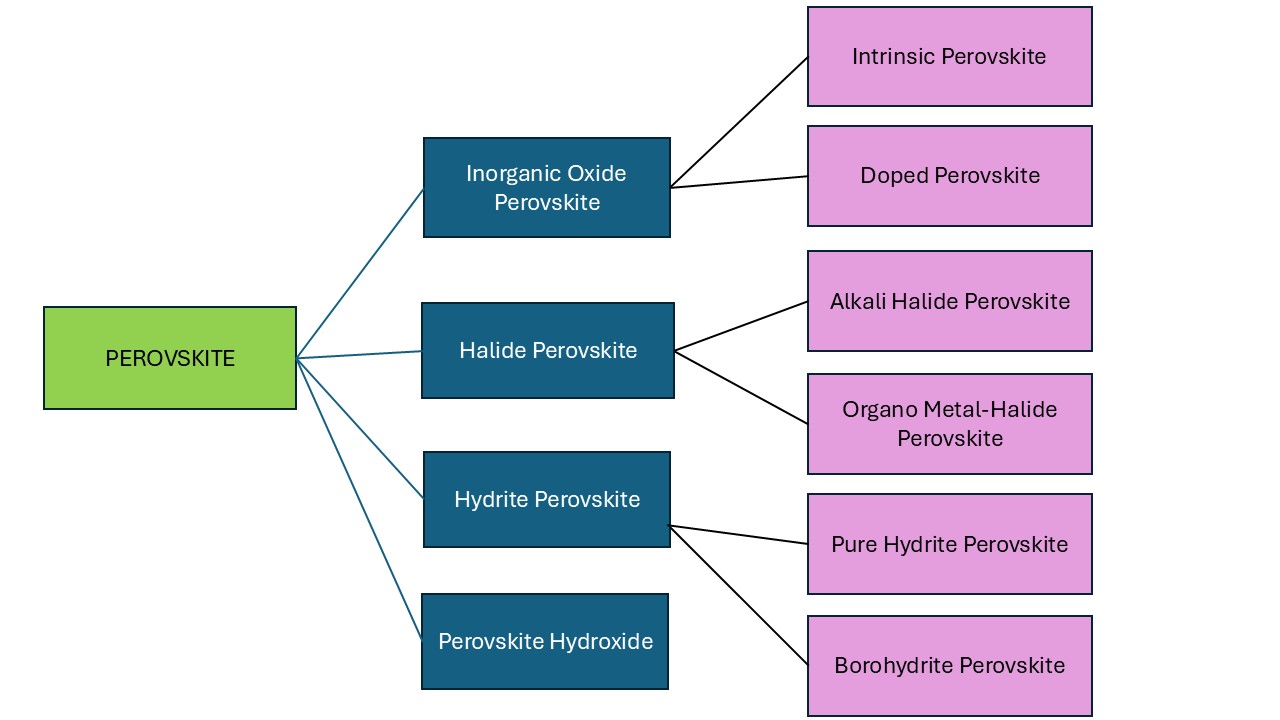}
        \caption{Family tree of Perovskite materials}
    \label{fig.8}
\end{figure}

Perovskites are considered to be one of the most valuable classes of materials from a technological point of view due to their simple chemical composition, structural stability, flexibility in fabrication, abundant availability on Earth's crust, multiple oxidation states, tolerance to doping and a wide range of distortions. So far, a large number of perovskite materials have been discovered and synthesized and categorized them in different families [Fig.\ref{fig.8}] The accommodation of such rich functional properties has made both single and double perovskites an acclaimed figure in various fields of research such as photovoltaics, solar cell\cite{CELESTINE2025}, catalysis\cite{Puia2026}, magnetism\cite{Lalhumhima2024}, ferroelectrics\cite{Kashikar2021}, spintronics\cite{Wei2021}, optoelectronics\cite{Chinggelkim2025}, etc. The manipulation of structural phases via doping, distortions, bonding mechanisms, and thermodynamic parameters is the key driving force; at the same time, room-temperature stability is crucial. Hence, understanding the crystal structure, compositions and symmetry under controlled structural manipulation plays a major role in optimizing the functional properties for desired technological applications.

\subsection{Single Perovskites}
Single perovskites have the general formula of ABX$_3$, where the A-cation typically occupies a large octahedral cavity, the B-cation sits at the centre of an octahedron, and the X-anion forms the coordinated octahedral network (see Fig.\ref{fig.9}). In the majority of cases, the ideal structure crystallizes in the cubic \(Pm\bar{3}m\) symmetry, in which perfectly aligned BX$_6$ octahedra form the corner-sharing 3D complex arrangement. The A cations are located in a 12-fold coordinated environment, holding the lattice using ionic interactions. However, the single perovskites do not only crystallizes in the ideal cubic phase; in many cases, the perovskite crystal structure deviates from perfect symmetry.\cite{Fu2022} The stability of the perovskite crystal (cubic structures) is measured from the Goldschmidt tolerance factor ($t$)\cite{Goldschmidt},
\begin{equation}
t = \frac{r_{A}+r_{X}}{\sqrt{2}(r_{B}+r_{X})}
\end{equation}
\begin{equation}
\mu=\frac{r_B}{r_X}
\end{equation}
where r$_{A}$, r$_{B}$, and r$_{X}$ denote the ionic radii of the A-cation, B-cation and X-anion. The value $t\approx 1$ signifies the most stable structure, and deviation from 1 leads to an unstable structure owing to structural distortions such as octahedral tilting, defects, ion displacement, and symmetry reduction (tetragonal, orthorhombic, rhombohedral, etc.). The $\mu$ values within 0.492-0.592 for the 3D structure of halide perovskite are an indication of a stable configuration.

Octahedral tilting is the most common form of distortion and is energetically favourable (see Fig.\ref{fig.9}a). The twisting of BX$_{6}$ octahedra modifies the B-X-B bond angles, eventually alters the overlapping mechanism between the electronic orbitals, and modulates the various physical properties such as magnetic, electronic, optical, transport, dielectric, etc. In materials like LaMnO$_{3}$ or CaTiO$_{3}$, the electronic bandwidth and magnetic super-exchange behaviour were determined by the tilting octahedral factor.\cite{Kim2009}

Another key distortion involves off-centering of the \(B\)-site cation, which breaks inversion symmetry and produces ferroelectricity (see Fig.\ref{fig.9}a). Classic examples include \(BaTiO_{3}\) and \(KNbO_{3}\), where the displacement of the central cation results in spontaneous electric polarization. Such distortions are highly sensitive to external stimuli including temperature, pressure, and electric fields, enabling tunable multifunctional behaviour.

Single perovskites also exhibit remarkable chemical flexibility. Substitutions at either the \(A\)- or \(B\)-sites allow tailoring of lattice parameters, electronic structure, charge compensation mechanisms, and defect chemistry. For instance, replacing the \(A\)-site cation with a smaller ion can enhance octahedral tilting, while doping the \(B\)-site can introduce new magnetic or catalytic functionalities. This tunability makes single perovskites a powerful platform for engineering materials with specific band gaps, magnetic ordering temperatures, or dielectric responses.

\subsection{Double Perovskites}
\begin{figure*}[h!]
    \centering
    \includegraphics[width=0.95\linewidth]{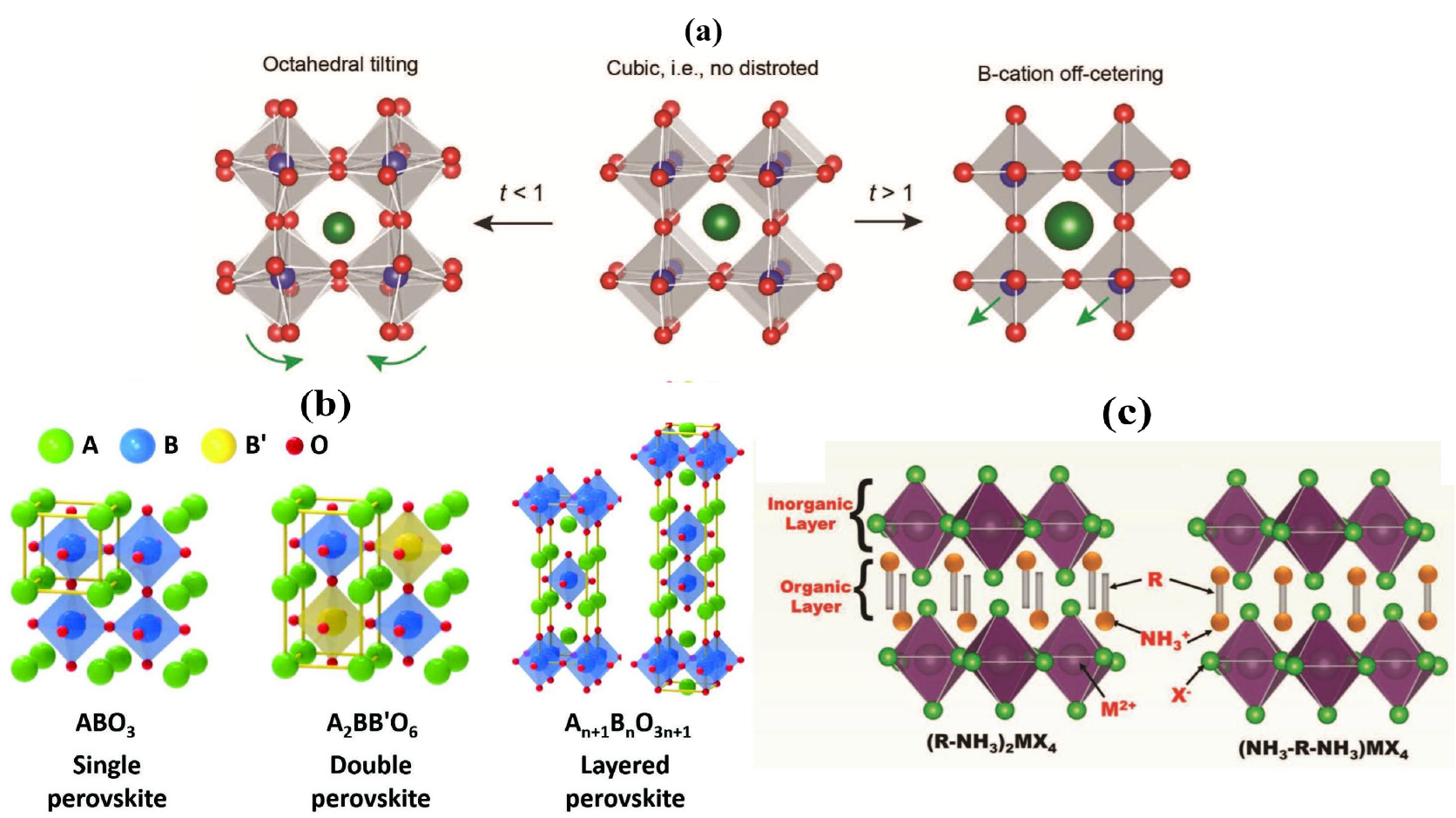}
    \caption{Different types of Perovskite materials: (a)Crystal structures of an ideal cubic perovskite and distorted perovskites: octahedral tilting and B-cation off-centering. Reproduced from \cite{Fu2022} with permission from John Wiley and Sons (b) Single, double, and layered perovskite crystal structure. Reproduced from \cite{Song2021} with permission from John Wiley and Sons (c) 2D layered hybrid organic-inorganic perovskite for n = 1 with mono and disubstituted amines.Reproduced from \cite{Krishna2019} with permission from John Wiley and Sons}
    \label{fig.9}
\end{figure*}

Double perovskites extend the structural diversity of the perovskite family through the ordering of two distinct \(B\)-site cations, resulting in materials with the general formula \(A_{2}BB'X_{6}\) (see Fig.\ref{fig.9}b).\cite{Song2021} These systems are often derived from the single perovskite structure through rock-salt-type ordering of the \(B\) and \(B'\) cations. When perfectly ordered, double perovskites commonly adopt the cubic \(Fm\bar{3}m\) structure, which features alternating \(BX_{6}\) and \(B'X_{6}\) octahedra.\cite{VASALA2015}

Cation ordering in double perovskites depends strongly on ionic radius differences, oxidation states, and electronegativity contrasts between the \(B\) and \(B'\) species. Larger differences promote complete ordering and enhance structural stability, while smaller differences can result in partial or complete disorder, lowering the symmetry to tetragonal, monoclinic, or orthorhombic variants.\cite{ANDERSON1993, KING2010}

The presence of two types of octahedra introduces more complex structural distortions compared to single perovskites. Octahedral tilting patterns become more intricate because the rotations of \(BX_{6}\) and \(B'X_{6}\) units may differ. Variations in bond lengths and angles between the two octahedral types significantly affect the electronic band structure, phonon modes, and magnetic interactions.\cite{Glazer1972}

Double perovskites exhibit unique electronic and magnetic behaviours not found in simpler perovskites. For example, materials such as \(Sr_{2}FeMoO_{6}\) display half-metallicity, where electrons of one spin channel exhibit metallic conductivity while the opposite spin channel remains insulating. This property is crucial for spintronic applications. Similarly, double perovskites containing heavy transition metals such as iridium or osmium show pronounced spin--orbit coupling, enabling exotic magnetic states and topological phenomena.\cite{Serrate2007, Witczak-Krempa2014, Cao2018}

Halide double perovskites have attracted interest for photovoltaic applications. Compounds like \(Cs_{2}AgBiBr_{6}\) demonstrate enhanced thermal and environmental stability compared to lead-based single halide perovskites. The ordered distribution of \(Ag^{+}\) and \(Bi^{3+}\) reduces defect formation and suppresses ion migration, resulting in improved structural robustness and prolonged operational lifetimes.\cite{McClure2016, Slavney2016}

The stability of perovskites, whether single or double, is influenced by geometric, electronic, and thermodynamic factors. In single perovskites, stability is primarily evaluated using the tolerance factor and the octahedral factor \(\mu = r_{B}/r_{X}\). Only certain combinations of ionic radii yield geometrically stable octahedra and an overall coherent framework. Strong covalency between the \(B\)-site cation and the anion, favourable charge balance, and the ability to accommodate strain through octahedral distortions also contribute to stability.\cite{Bartel2019, Li2017}

Double perovskites require an additional criterion related to \(B/B'\) cation compatibility. Charge balance, oxidation-state combinations, and ionic-size mismatch play critical roles in determining whether the structure is stable and whether cation ordering is favoured. Ordered double perovskites are often more stable because alternating octahedra relieve lattice strain and create energetically favourable environments. Entropic contributions may stabilize disordered phases at high temperatures.\cite{Yu2019}

Defect tolerance is another key aspect of perovskite stability. Single perovskites can accommodate oxygen vacancies, cation substitutions, and local distortions with minimal structural disruption, making them robust for catalytic and ionic conduction applications. Double perovskites, particularly the halide variants, often exhibit higher resistance to thermal decomposition and photodegradation due to their strong cation-ordering tendencies and reduced defect mobility.\cite{Brandt2017}

Single and double perovskites constitute a structurally rich and highly adaptable family of materials whose properties arise from the interplay of octahedral geometry, cation arrangement, and lattice distortions. While single perovskites offer straightforward tunability through simple ABO$_{3}$ frameworks, double perovskites introduce additional structural complexity through ordered substitution at the \(B\)-site, enabling exotic electronic and magnetic behaviour (see Table \ref{Table 1}). Understanding the structural properties and stability of both classes is fundamental for advancing their applications in energy conversion, electronics, spintronics, catalysis, and beyond.\cite{VASALA2015, McClure2016}

\begin{table*}[h!]
\centering
\renewcommand{\arraystretch}{1.5}
\begin{tabular}{p{5cm} p{5cm} p{6cm}}
\hline
\textbf{Feature} & \textbf{Single Perovskite} & \textbf{Double Perovskite} \\
\hline
Formula & $\mathrm{ABX_3}$ & $\mathrm{A_2BB'X_6}$ \\
\hline
Structure Type & Corner-sharing $\mathrm{BX_6}$ octahedra & Alternating $\mathrm{BX_6}$ and $\mathrm{B'X_6}$ octahedra \\

Common Symmetry & Cubic, tetragonal, rhombohedral, orthorhombic & Cubic, tetragonal, monoclinic \\
Stability Criterion & Goldschmidt tolerance factor & Extended tolerance factor using average B-site radius \\
Octahedral Distortions & Tilting, Jahn--Teller distortion, cation displacement & Breathing modes, multi-octahedral tilting, B/B' ordering \\
Example Compounds & $\mathrm{BaTiO_3}$, $\mathrm{SrTiO_3}$, $\mathrm{LaMnO_3}$ & $\mathrm{Sr_2FeMoO_6}$, $\mathrm{Cs_2AgBiBr_6}$ \\
Key Properties & Ferroelectricity, magnetism, electrical conductivity & Half-metallicity, spintronics, enhanced photostability \\
Defect Tolerance & Moderate to high & High (especially HDPs) \\
Applications & Dielectrics, catalysis, PV, energy-storage, thermoelectrics, piezoelectric, photonics, sensors & Spintronics, optoelectronics, quantum materials, energy-storage, thermoelectrics, piezoelectric, photonics, sensors \\
\hline
\end{tabular}
\caption{Comparison of structural features and stability characteristics of single and double perovskites with key references. \cite{Goldschmidt, ANDERSON1993, VASALA2015, McClure2016}}
\label{Table 1}
\end{table*}

\section{Electronic and Optical Properties}
Perovskite materials are renowned for their exceptional electronic characteristics, including tunable direct band gaps, high carrier mobility, and defect tolerance. Direct band gaps are often observed in perovskite materials, which is an important factor in both light absorption and the formation of charge carriers. Changing the composition of a material by replacing an anion or cation with another element can either widen or reduce the band gap. For example, MAPbI$_3$ possessing a band gap of 1.55 eV increases to 2.23 eV when using bromine instead of iodine.\cite{RATURI2026} A high charge carrier mobility is crucial for efficient charge transport through a solar cell. The electron and hole mobilities of perovskite fall in the range of 10-100 cm$^2$/V.s. This can vary according to the composition as well as the processing environment of the compound.\cite{Chen2019} The high carrier mobilities can reduce recombination losses and enhance the PCE. Further, when compared to other semiconducting materials, perovskites are less prone to defects. Their electronic properties are not significantly affected by defects and grain boundaries, thus achieving high-performance solar cells even with an inexpensive solution processing method.\cite{Singh2020}

\par Understanding the fundamental optical properties such as dielectric constant, absorption coefficient, and refractive index is critical to utilize in solar cells. The high static dielectric constant of perovskite allows for lower binding energy of electron-hole pairs while also shielding charge carriers from defects.\cite{SHAH2025} Their absorption coefficient can reach up to 10$^5$ cm$^{-1}$ in the visible region, which is much higher than crystalline silicon, allowing for a thinner active layer ($\sim$400 nm) in solar cell devices.\cite{CHEN2026} This is attributed to their direct band gap plus high refractive index causing efficient light absorption and conversion to charge carriers. The refractive index is an important factor that indicates light interaction with the material and determines its reflection and refraction behaviour to decide whether the material is applicable in optoelectronic devices.\cite{RATURI2026} Perovskite materials typically exhibit high refractive indices; therefore, a specific anti-reflection coating is necessary to minimize reflection losses. 

\section{Comparative Analysis} The next-generation solar technology is rapidly developing to use lead-free double perovskites. They provide us with the stability and low toxicity that we require, as well as high optoelectronic potential. Their practical effectiveness is not yet up to lead-based perovskites, but with numerical simulations, we can bridge that divide. The tools, such as SCAPS-1D, are now needed to test newer architectures and determine what exactly is holding them back. This section provides an in-depth exploration of these systems, which have significant potential, all discussed using SCAPS-1D optical simulations. We subdivide the analysis into four main points: Bandgap analysis of all absorbers, role of Electron Transport Layers (ETLs), role of absorber properties, and the role of Hole Transport Layers (HTLs).
\subsection{\textbf{Bandgap Analysis}} 
\begin{figure}[h!]
    \centering
    \includegraphics[width=0.8\linewidth]{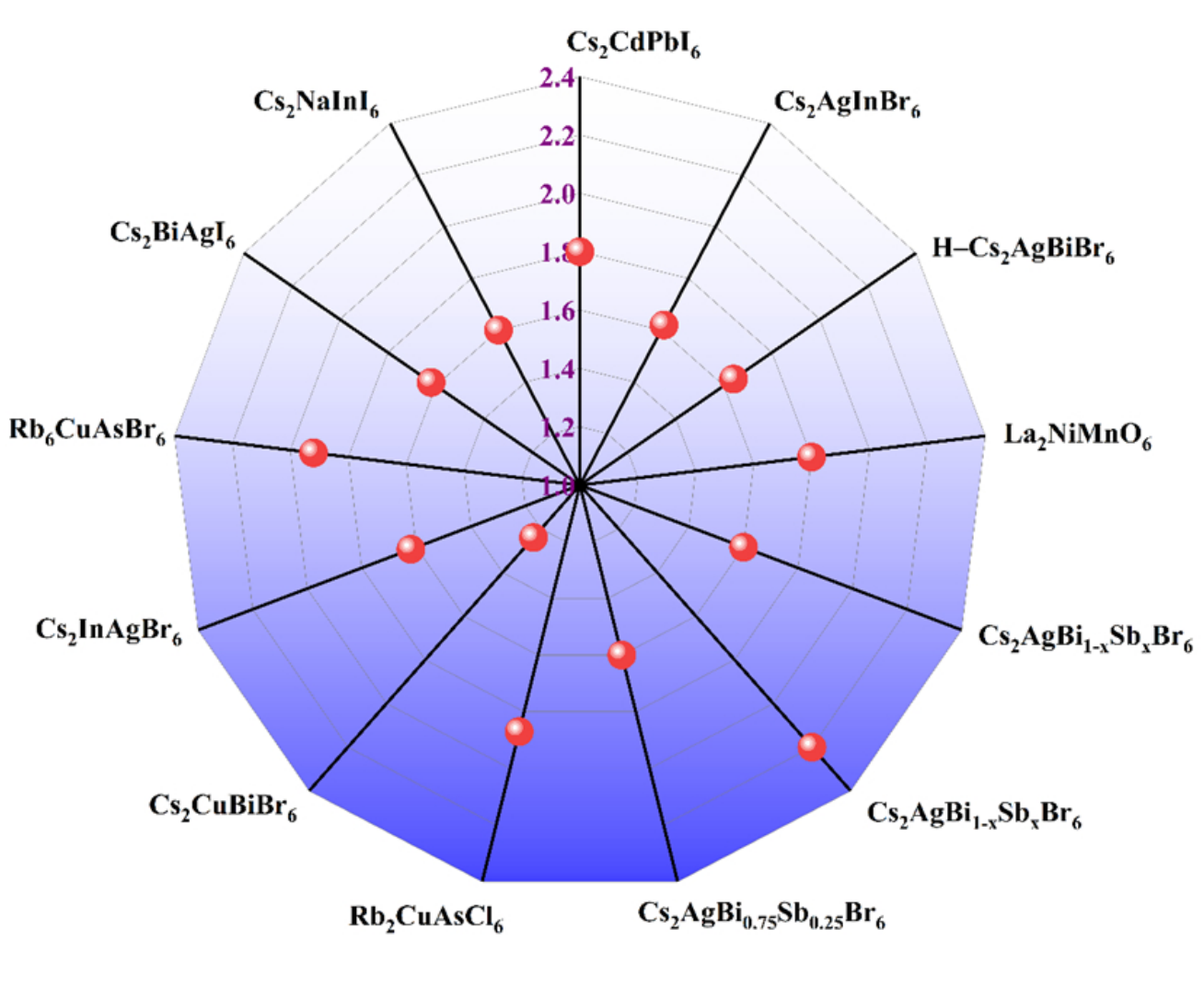}
    \caption{Bandgap Analysis of various double perovskites}
    \label{fig.10}
\end{figure}

One of the key parameters used to measure the performance of both theoretical and simulated solar cells of the type of double-perovskites is the electronic band gap. Fig.\ref{fig.10} shows band gap analysis of various double perovskites. Working with SCAPS simulations, the band gap itself has direct effects on the critical device parameters, such as the absorption coefficient, the intrinsic carrier concentration, the open-circuit voltage (V$_{OC}$) and the spatial peculiarities of the depletion region.\cite{Ullah2024} In the analysis of the eleven absorbers in this research, we find that there is a wide range of band gaps, with narrow band gaps in the range of 1.24 eV up to broader gaps with values of over 1.80 eV. The difference produces a heterogeneous terrain of possible current generation and voltage outputs. At the smaller extreme, Cs$_2$CuBiBr$_6$ has a band gap of $\sim$ 1.24 eV.\cite{Utsho2025} Simulations show that this lowered gap improves near-infrared absorption, giving a significant increase in short-circuit current (J$_{SC}$). Though this inherently restricts the V$_{OC}$, the wider absorption range enables the material to reach simulated efficiencies of 19-20 if the transport layers are well-aligned. On the other hand, Rb$_2$CuAsCl$_6$ has a significantly larger band gap of about 1.87 eV.\cite{Assiouan2025} Whereas wide band gaps generally limit long-wavelength absorption, the material has broken this rule; it has large absorption coefficients due to its own DFT inputs,\cite{SHAHZAD2024}, and it provides an impressive J$_{SC}$ of almost 38 mA/cm$^2$. Contrary to this, its bromide equivalent, Rb$_2$CuAsBr$_6$, exhibits poorer optical absorption and thus leads to poor performance. Going beyond the typical halide structure, La$_2$NiMnO$_6$ (LNMO) is an oxide double perovskite structure that has a mid-gap value of 1.8 eV.\cite{Singh2023} Although this gap is much broader, the inherent electronic structure of LNMO provides efficient charge separation and long carrier lifetimes to achieve high J$_{SC}$ values ($\sim$28 mA/cm$^2$) and a Power Conversion Efficiency (PCE) of about 25.4\%. Also, considerable enhancements are possible if the band gap is chemically modified,\cite{Miah2024}, i.e. hydrogenating Cs$_2$AgBiBr$_6$ decreases the band gap of this substance to 1.64 eV. This change significantly enhances the visible-light absorption and electronic compatibility with transport layers, which leads to a 27.3\% PCE.\cite{Sabbah2024} Equally, the band gap is reduced when the Cs$_2$AgBiBr$_6$ is alloyed with antimony to give Cs$_2$AgBi$_{0.75}$Sb$_{0.25}$Br$_6$ to enhance the spectral response, and produce a PCE of 22\%.\cite{Asif2025} This is also observed in the wider family of Cs$_2$AgBi$_{1-x}$Sb$_x$Br$_6$, where simulations find the optimum balance between current and voltage to be x = 0.25, which is refracted too narrowly to allow adverse recombination losses. Several materials naturally fall within the optimal photovoltaic window of 1.5–1.8 eV without the need for modification.\cite{Hossen2025} Cs$_2$InAgBr$_6$ has a direct band gap of 1.62 eV, which favours better optical absorption and a simulated PCE of 26.64\% when coupled with ZnSe and MASnBr$_3$.\cite{Abrar2025} Likewise, Cs$_2$CdPbI$_6$ capitalises on its 1.8 eV gap and high absorption coefficients to achieve one of the highest theoretical efficiencies in the experiment ($>$32\%), demonstrating the possibility of mixed-metal systems.\cite{Mohamed2025} Cs$_2$NaInI$_6$ also demonstrates an excellent balance with a direct gap of 1.60 eV; its good compatibility with high-mobility ETLs, including WS$_2$, makes it possible to use in the case of PCE of 22.6\%.\cite{Rahman2025} Though Cs$_2$InAgBr$_6$ has a similar bandgap (1.62 eV), the effective masses are different, causing a slightly lower PCE of 19.26\%.\cite{Abrar2025} Lastly, in tandem design, Cs$_2$BiAgI$_6$ (1.7 eV) is effective in harvesting long-wavelength light.\cite{Prasana2024} As a final point, it is important to note that simulations of SCAPS show that the sweet point of a highly efficient double perovskite is between 1.5 and 1.8 eV, which can be visualized in the comparative radar chart (Fig.\ref{fig.10}), indicating that the structural tunability of the A$_2$BB’X$_6$ family is enormous.

\subsection{\textbf{Electron Transport Layers (ETL) Effect on the performance of the device}} Determining the carrier extraction efficiency, V$_{OC}$, J$_{SC}$, and overall recombination dynamics of the double-perovskite solar cells is determined by the choice of the ETL.\cite{SOUDAGAR2025} It is clear that the choice of the optimal ETL is highly constrained in a strict sense by physical parameters: alignment of the conduction bands, interface defect density, carrier mobility, and doping concentration. The material constituting a minimal or slightly positive conduction-band offset (CBO) with the absorber produces the best performance. This particular orientation forms an energetic scenario, causing an easy extraction of the electrons and a good prevention of the back-injection of holes.
\subsubsection{\textbf{Optimised Band Alignment and High Efficiency}} Of the particular combinations considered, ZnSe would be one of the most promising candidates to be used in Cs$_2$AgInBr$_6$ devices since it has virtually perfect conduction-band matching. The simulations validate the assertion that a near-zero CBO has a strong decreasing effect on the recombination rates at the ETL/absorber interface,\cite{SINHA2022}, which is the direct cause of the device having one of the highest simulated efficiencies in this study ($\sim$26.64\%). Likewise, ZnO is the best alternative to hydrogenated Cs$_2$AgBiBr$_6$. It has a good band level and low intrinsic defect density that contributes to the easy extraction of the electrons, which promotes a PCE of 26.3\%.
\par There were also large improvements in gains by substituting TiO$_2$ with SnO$_2$ in Cs$_2$CdPbI$_6$ devices.\cite{Medjaldi2020} According to the simulations, it shows that SnO$_2$ stabilizes the depletion region, as well as increases V$_{OC}$ by a reduced recombination coefficient, driving the efficiency to more than 32\%. Moreover, ZnOS is of great use with La$_2$NiMnO$_6$ (LNMO) and the Cs$_2$AgBi(Sb)Br$_6$ alloy system. In such instances, ZnOS is used to increase the inherent potential and expand the depletion region to increase the collection efficiency and drive PCEs to over 25\% and 23\%, respectively.
\subsubsection{\textbf{The Role of \texorpdfstring{WS$_2$}{WS2} and the "Spike" Structure}} WS$_2$ was found on several occasions to be a high-performance ETL, especially towards Cs$_2$NaInI$_6$, Cs$_2$CuBiBr$_6$, and Cs$_2$InAgBr$_6$.\cite{Utsho2025, Rahman2025, Abrar2025} The main reason why it is successful is because of its high electron mobility and the generation of a favourable spike alignment. Through the use of a Cs$_2$NaInI$_6$ layer, as an example, WS$_2$ forms a small positive CBO to form an energetic barrier to hole back-flow without impeding electron transport. This process enables the device to achieve a PCE of 22.6\% at a high current density of 21.4 mA/cm$_2$. As SnS$_2$ provides a competitive performance owing to comparable alignment, other substances such as In$_2$S$_3$ provide a conduction-band cliff, which adversely raises interfacial recombination.\cite{TRIPATHI2021} On the same note, though IGZO is also favourable to fabrication, its amorphous structure is sensitive to defect states, resulting in the deterioration of efficiency.\cite{LEE2025}
\subsubsection{\textbf{Limitations of Standard Oxides}} Although TiO$_2$ is still considered a conventional ETL with Rb-based absorbers (Rb$_2$CuAsBr$_6$ and Rb$_2$CuAsCl$_6$) as they offer a convenient barrier and high internal electric field (PCE = 20.6\%), a serious weakness has been identified in their analysis: they are sensitive to defects. When the density of ETL defects goes beyond 10$^{16}$ cm$^{-3}$, the performance of TiO$_2$-based devices degrades substantially, indicating that with these double perovskites, the quality of the ETL is no less important than the material choice itself.\cite{Bulowski2024} Taken together, these SCAPS outcomes indicate a ranking of ETL efficacy of double perovskites. Mobile materials, which allow focal band engineering, namely WS$_2$, ZnSe, SnO$_2$, ZnO, and ZnOS, are by far better than the traditional ones, such as TiO$_2$ or organic ones (C$_{60}$, PCBM), especially when there is a minimal number of interface defects.
\subsection{\textbf{Effect of Absorber Properties on Device Performance}} The most vital of the components that determine the simulated performance of the double-perovskite solar cells is the absorber layer.\cite{CHAUHAN2023} SCAPS-1D can be used to critically analyse several absorber parameters, such as the thickness of the layers, defect density, interface quality, doping concentration, band structure, and recombination, reasonably. In all eleven absorbers that have been studied in this paper, there are general trends that are consistent to the point of directly relating the physical and electronic properties of the absorber to the photovoltaic measures that they give.
\subsubsection{\textbf{Bandgap and Intrinsic Properties of Double Perovskites}} In Figure \ref{fig.10}, a radar-type comparison is shown that supports the high dependence of the performance of the devices on the intrinsic optical properties of the absorber. The broad chemical tunability of A$_2$BB’X$_6$ double perovskites is indicated by the wide bandgap distribution, which spans between about 1.24 eV and more than 2.2 eV. Narrow-bandgap materials, like Cs$_2$CuBiBr$_6$, prefer near-infrared absorption much higher, which is a powerful stimulator of J$_{SC}$ but a strong inhibitor of V$_{OC}$. Mid-band gaps, on the other hand, such as Cs$_2$AgInBr$_6$, hydrogenated Cs$_2$AgBiBr6, Cs$_2$NaInI6 and Sb-alloyed Cs$_2$AgBi$_{1-x}$Sb$_x$Br$_6$ are located in the optimum photovoltaic window (1.5-1.7 eV) and therefore have the best current-voltage ratio. The wider-bandgap absorbers, such as La$_2$NiMnO$_6$, Cs$_2$CdPbI$_6$ and Rb$_2$CuAsCl$_6$, are chiefly better V$_{OC}$, but require a trade-off to be less active at long wavelengths.
\subsubsection{\textbf{Thickness, Defects, and Doping Analysis}} There is a direct relationship between absorber thickness and carrier generation and photon absorption. The optimum thickness of most materials is conservatively between 500 and 900 nm.\cite{Rozhko2021} Indicatively, Cs$_2$AgInBr$_6$ is optimised at a wavelength of about 600 nm, where J$_{SC}$ of over 27 mA/cm$^2$ is realised. Narrower-bandgap materials, such as Cs$_2$CuBiBr$_6$, and those with more active absorption, such as hydrogenated Cs$_2$AgBiBr$_6$, demand a somewhat thicker layer, 700 nm, to be able to take full advantage of their absorption capability. It is also worth noting that Cs$_2$NaInI$_6$ would need a relatively thicker layer, and the performance would be improved till 1200 nm (J$_{SC}$ 22.9 mA/cm$^2$) because of better photon absorption and lower transmission losses. Nevertheless, for all materials, the thickness that is beyond the optimum range stretches the transport paths, which enhances bulk recombination and consequently lowers the FF and  V$_{OC}$.\cite{IKUEMONISAN2025} Defect density (N$_t$) of absorbers is also another very influential factor.\cite{LEKSHMY2025} The SCAPS simulations clearly indicate that N$_t$ should not exceed 10$^{15}$ cm$^{-3}$ in all materials to maintain high V$_{OC}$ and FF.\cite{Zeghdar2025} Cs$_2$CdPbI$_6$ and Sb-alloyed Cs$_2$AgBi$_{1-x}$Sb$_x$Br$_6$ exhibit acute deep trap sensitivity, and the carrier lifetime is sharply reduced, leading to severe V$_{OC}$ collapse. Likewise, Rb$_2$CuAsCl$_6$ is readily degraded in J$_{SC}$ with N$_t$ greater than 10$^{15}$ cm$^{-3}$, which shows that defects must be strongly prevented in the bulk absorber. Similarly, defects in the interface are also harmful.\cite{Stanley2022} In the case of materials such as Cs$_2$AgInBr$_6$ and Cs$_2$AgBi$_{0.75}$Sb$_{0.25}$Br$_6$, it is important to reduce interface traps because it lengthens the depletion region, enhances charge separation, and ensures a high FF and J$_{SC}$.\cite{NJEMA2024} 
\par In terms of doping concentration, moderate p- or n-type doping reinforces the intrinsic electric field that can result in charge extraction, as is observed in LNMO and Cs$_2$AgBi$_{1-x}$Sb$_x$Br$_6$. Nevertheless, doping should not be excessive, as it can lead to harmful Auger recombination and undermine the device's long-term stability.\cite{Lu2024}
\subsubsection{\textbf{Composition Tuning and Performance Synthesis}} Another strategy which is effective in high-performance optimization is absorber composition tuning.\cite{Nowak2021} The incorporation of Sb into Cs$_2$AgBiBr$_6$ achieves the reduction of the band gap and the increase of the absorption in the visible range. Other positive improvements are also realized by the incorporation of hydrogen in the same host material.\cite{BORETTI2025} On the other hand, although Cs$_2$BiAgI$_6$ is limited in potential in a single-junction device, its band structure is very efficacious as a lower-cell absorber in tandem constructions. The combined effect of these absorber-related factors is reflected in the simulated PCE results. 
\par
\begin{figure}[h!]
  \centering
    \includegraphics[width=0.8\linewidth]{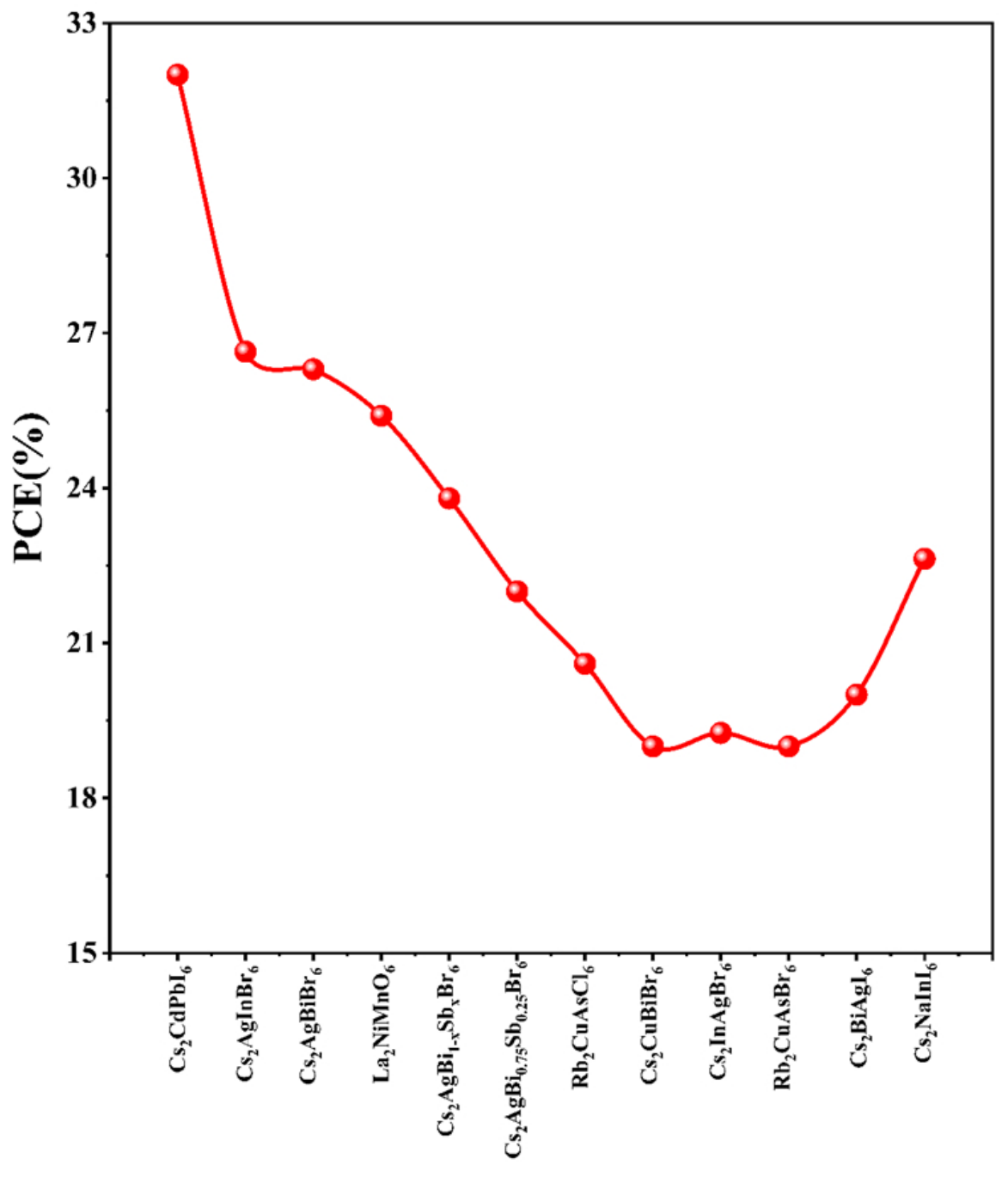}
    \caption{Band-gap and performance comparison of double perovskite absorber}
   \label{fig.11}
\end{figure}

 Figure \ref{fig.11} makes a comparison of the simulated PCEs of the studied materials. With the most favourable PCEs ranging between 25\% and 32\%, these include Cs$_2$CdPbI$_6$, Cs$_2$AgInBr$_6$, hydrogenated Cs$_2$AgBiBr$_6$, LNMO and the Sb-alloyed compositions. Their extreme efficiencies can directly be attributed to the best band gaps, high absorption coefficients and good electronic characteristics. Mid-range materials (Cs$_2$AgBi$_{0.75}$Sb$_{0.25}$Br$_6$, Cs$_2$AgBiBr$_6$ and Rb$_2$CuAsCl$_6$) obtain PCEs of 20-23\%, constrained by moderate band gaps or increased trap densities. Reduced efficiencies (18.20\%) are seen with materials such as Cs$_2$CuBiBr$_6$, Cs$_2$InAgBr$_6$ and Cs$_2$BiAgI$_6$, either because of high recombination or undesirable optical transitions. Cs$_2$NaInI$_6$, having a PCE of 22.6\%, has been shown to exhibit high and steady performance in the case when defect density and thickness have been optimised. In sum, absorber optimisation is needed to achieve the theoretical performance limits of the SCAPS-1D light absorption ability. Defects, interface passivation, doping, and bandgap engineering are always found to be the most influential factors.\cite{Banik2024}
\subsection{\textbf{Impact of the Hole Transport Layer (HTL)}} The Hole Transport Layer (HTL) is just as important as the ETL and serves to remove generated holes, which is the main purpose of the solar cell (prevention of electron back-flow and creation of energy selectivity at the rear junction of the solar cell as well as creating energy selectivity). Our multi-perovskite wide-scale SCAPS-1D modelling experiments demonstrate that HTL selection also has a dominant influence on V$_{OC}$, FF, series resistance, and, what is most important, HTL/absorber interface recombination.
\subsubsection{\textbf{Inorganic HTLs and Optimal Alignment}} Inorganic materials were recurring as highly efficient HTLs with the required stability and high band alignment.\cite{ZHOU2026} CuI was particularly successful, particularly when Cs$_2$CdPbI$_6$ and Cs$_2$InAgBr$_6$ were concerned. CuI provides good valence-band alignment and high hole mobility in the Cs$_2$CdPbI$_6$ system, which is the main factor behind the simulated PCE of over 32\%.\cite{Ghosh2024} Likewise, its application to Cs$_2$InAgBr$_6$ allows the extraction of holes to be extracted effectively, back-interface recombination to be reduced and FF to be greatly improved. CuSCN proved to be even better when used together with Rb$_2$CuAsCl$_6$, Rb$_2$CuAsBr$_6$ and Cs$_2$AgBi$_{0.75}$Sb$_{0.25}$Br$_6$ absorbers. SCAPS modelling has revealed that the valence band of CuSCN would be slightly higher than that of these absorbers. This slight offset permits easy transport of holes and, at the same time, high energy selectivity, which leads to higher FF, better V$_{OC}$ and stability in general.\cite{Mercy2024}
\par In the case of Cs$_2$AgInBr$_6$, the optimum HTL was identified as MASnBr$_3$, which was best suited with respect to position in the valence band and hole conductivity. Simulations show that MASnBr$_3$ forms a significantly higher J$_{SC}$ and FF in comparison with other materials such as Cu$_2$O and CBTS. More so, Cu$_2$O is the material of choice in LNMO-based devices; its ideal valence-band alignment has led to recombination being minimized, with the device having an impressive FF of over 82\% and a PCE of 25.4\%. Even in the narrow-bandgap Cs$_2$CuBiBr$_6$, CBTS was superior to CuSCN, mostly because of the advantage of better band alignment, that is, essentially a reduction of series resistance as well as a high FF at different simulated absorber thicknesses.
\subsubsection{\textbf{Organic HTLs and Limitations}} On the contrary, organic HTLs, including Spiro-OMeTAD, although occasionally exhibiting high J$_{SC}$ and V$_{OC}$ (as observed with hydrogenated Cs$_2$AgBiBr$_6$), brought about significant instability and sensitivity to interface traps and substrate thickness. Though this organic material can temporarily raise the performance parameters, due to its inherent instability, it cannot be used in the long term, especially when compared to the high-performance inorganics.\cite{ALANAZI2025} Finalisation of HTL requirements together, the simulations of the SCAPS have remained consistent on the fact that the best HTLs are those that offer the following three attributes: low defect density, good valence-band alignment and high hole mobility.\cite{ELBADRAOUI2026} CuI, CuSCN, MASnBr$_3$, CBTS, and Cu$_2$O all consistently perform well in the entire double-perovskite systems studied because they have positive energy-level alignment as well as strong electrical characteristics.
\subsubsection{\textbf{Comparative Summary Tables}} This part is a synthesis of the major simulation outcomes, providing a comparative summary table of the eleven double-perovskite absorber systems simulated with the help of the SCAPS-1D platform (see Table \ref{Table 2}). These tables are a summary of the intrinsic material properties (band gap)and the simulated photovoltaic performance (PCE) discovered within the scope of the study.
\begin{table*}
\small
  \caption{Band gap and Peak SCAPS Efficiency. This table orders the eleven double-perovskite absorbers mainly by their maximum simulated Power Conversion Efficiency (PCE) to show the close relationship between the optimal bandgap and the overall device potential.}
  \label{Table 2}
  \begin{tabular*}{\textwidth}{@{\extracolsep{\fill}}llll}
    \hline
    Absorber material & Band gap (eV) & Peak PCE (\%) & \\
    \hline
    Cs$_2$CdPbI$_6$  & 1.8 & 32\cite{Mohamed2025} & \\
    Cs$_2$AgBiBr$_6$ & 1.64& 26.3\cite{Sabbah2024} &\\
    La$_2$NiMnO$_6$ & 1.8 & 25.4\cite{Singh2023} &\\
    Cs$_2$AgBi$_{1-x}$Sb$_x$Br$_6$ & 1.6-2.2& 23.8\cite{Hossen2025} &\\
    Cs$_2$NaInI$_6$ & 1.6& 22.63\cite{Rahman2025} &\\
    Cs$_2$AgBi$_{0.75}$Sb$_{0.25}$Br$_6$ & 1.60& 22\cite{Asif2025} &\\
    Rb$_2$CuAsCl$_6$ & 1.87& 20.6\cite{Assiouan2025}&\\
    Cs$_2$CuBiBr$_6$ & 1.24& 19-20\cite{Utsho2025} &\\
    Cs$_2$InAgBr$_6$ & 1.62& 19.26\cite{Abrar2025} &\\
    Rb$_2$CuAsBr$_6$ & 1.92& 18-19\cite{Assiouan2025} &\\
    Cs$_2$BiAgI$_6$ & 1.62& 20-22 (tandem)\cite{Prasana2024} &\\
    \hline
  \end{tabular*}
\end{table*}

Different Papers Reported PCE under different conditions, like lab-scale, stabilized, and certified values. Many studies focused on initial PCE rather than lifetime performance, which makes something like comparing their results, like durability across technologies, difficult. Some Data and results we get are from different sources and reports, and could not be obtained from a single paper. The rapid development of perovskite research poses some challenges while reviewing this paper since perovskite solar cells improve very quickly. 

\subsection {\textbf{Computational Methods}}
\subsubsection{Density Functional Theory (DFT)}
First-principles calculations were performed using Density Functional Theory (DFT) to obtain electronic structures, band gaps, and optical responses of double perovskite compounds. DFT plays a central role in understanding the optoelectronic physics of lead-free double perovskites such as \(Cs_2AgBiBr_6\). While experimental data (XRD, absorption, PL, PLE) describe how the material behaves, DFT reveals why it behaves that way by computing the fundamental electronic structure.\cite{Jones2015}

\subsubsection{DFT before SCAPS}
To effectively simulate the photovoltaic performance of double perovskites using SCAPS-1D, it is a must to gather a comprehensive set of intrinsic material parameters first, as SCAPS does not inherently compute these properties. Instead, we must input values like band gap energy, electron affinity, effective density of states, dielectric constant, carrier mobilities, absorption coefficient, and defect energy levels. Assuming or guessing these values could result in inaccurate or non-physical device performance predictions. Therefore, before employing SCAPS for device-level simulations, it is necessary to determine these parameters through a density functional theory (DFT)–based computational approach. DFT software, such as VASP, Quantum ESPRESSO, or CASTEP, enables the computation of electronic band structure, projected density of states (pDOS), effective masses of electrons and holes, optical constants (n, k), and dielectric response directly from the atomic structure of the material by using first principles.

\subsubsection{SCAPS-1D} 
It is a one-dimensional Solar Cell Capacitance Simulator and is now one of the most commonly used tools for modelling photovoltaic devices. With the rise of highly tunable perovskite materials, however, many studies have begun to report device efficiencies that are far above realistic values—some even exceeding the Shockley–Queisser limit. Although such results may appear promising, they often stem from nonphysical simulation parameters, including unrealistically low defect densities, extremely small radiative recombination rates, or excessively high doping levels. A review of more than 250 perovskite solar cell simulations shows that these parameter choices, along with methodological inconsistencies and the inherent optical limitations of SCAPS-1D, are the main causes of these exaggerated predictions. Therefore, the study recommends establishing standardized simulation practices. Overall, SCAPS outputs must be interpreted carefully, as the accuracy of the results depends strongly on the reliability of the input parameters and the software’s numerical approximations.\cite{SAIDARSAN2025}

To put it simply, DFT is used to compute key electronic parameters of double perovskite, such as band gap (Eg), band structure (direct/indirect), density of states (DOS), effective masses, dielectric constant, electron affinity, ionization potential/work function, absorption coefficient, and defect formation energies. These are the materials input for SCAPS-1D. After obtaining the parameters, SCAPS-1D is used to simulate the solar cell to obtain the I-V characteristics: the J–V curve, open-circuit voltage (Voc), short-circuit current (Jsc), fill factor (FF), power conversion efficiency (PCE), etc. 
In order to carry out the survey, keywords "SCAPS" and "Perovskite" were searched in respected journals, gathering data on device parameters such as Voc, Jsc, FF, and efficiency. These findings were compared with the top experimentally confirmed values from the Perovskite Database Project (PDP) as well as the Shockley–Queisser limits under AM1.5G illumination. The most frequently simulated absorber materials, MAPbI$_3$ and MASnI$_3$, were scrutinized along with their typical transport layers, Spiro-OMeTAD (HTL) and TiO$_3$ (ETL). The review aimed to pinpoint discrepancies in input parameters and optimization techniques. Furthermore, the role of unrealistic assumptions and the optical limitations of SCAPS-1D was assessed by simulating a reference device using its documented parameters and the default AM 1.5G spectrum in SCAPS-1D (v3.3.11).\cite{VERSCHRAEGEN20076276,BURGELMAN2000527}

\section{Limitations \& Challenges} Despite remarkable progress in simulating lead-free double perovskite solar cells using SCAPS-1D, several profound limitations and challenges persist across material properties, device physics, computational methodologies, and practical commercialization.
\par The most fundamental challenge lies in the inherent optoelectronic properties of double perovskite absorbers. Many lead-free double perovskites, such as Cs$_2$AgBiBr$_6$ and Cs$_2$AgBiCl$_6$, possess wide and indirect electronic band gaps (typically 1.8-2.3 eV), which lie outside the ideal Shockley-Queisser range for single-junction solar cells (1.1-1.5 eV). This mismatch fundamentally limits photocurrent and power conversion efficiency (PCE). While alloying strategies (e.g., Sb substitution) can narrow the band gap, they often introduce new defect states. Recent investigations of similar halide double\cite{ALAWAIDEH2026} confirm that, despite promising optoelectronic properties, challenges in band gap engineering and carrier mobility optimization persist. Many double perovskites exhibit large effective masses for electrons and holes, leading to low carrier mobility, poor charge extraction, and increased bulk recombination.
\par Double perovskites also introduce stability-conductivity trade-offs. Substituting  Ag$^+$ with cations like  In$^+$  or  Tl$^+$ can produce favourable direct band gaps but often destabilizes the crystal structure. Iodide-based double perovskites offer the best band gaps but are notoriously unstable, while stable chloride and bromide variants have suboptimal optical properties. Oxidation sensitivity of B-site cations like  Sn$^{2+}$  or  Ge$^{2+}$ remains severe, with rapid oxidation creating deep-level defects. Comprehensive DFT studies on related materials\cite{ALAWAIDEH2025} reveal that structural stability and electronic properties are highly composition-dependent, with different perovskite families showing varied degradation mechanisms, highlighting the need for material-specific assessments.
\par SCAPS-1D simulations consistently highlight extreme sensitivity to defects. Absorber bulk defect densities (Nt) must remain below 10$^{15}$cm$^{-3}$ to achieve high efficiencies, a challenging target experimentally. Interface recombination at ETL/absorber and HTL/absorber junctions represents the dominant loss mechanism, with misaligned band offsets creating pathways for non-radiative recombination described by the Shockley-Read-Hall model. Investigations of related ternary compounds\cite{ALAWAIDEH20261} demonstrate how comparative analysis of structural and electronic properties provides insights into defect formation energies. Similarly, studies on MAX phases\cite{Rached2026} emphasize the importance of verifying stability through energy calculations before recommending materials for applications.
\par Selecting appropriate charge transport layers is fraught with constraints. Materials like WS$_2$, ZnSe, and SnO$_2$ show high simulated performance, but experimental integration requires near-perfect band alignment. Standard oxides like TiO$_2$ exhibit low electron mobility (0.1-4 cm$^2$/Vs), creating bottlenecks for charge extraction. Organic HTLs like Spiro-OMeTAD introduce instability due to their hygroscopic nature and require dopants. Innovative interface engineering approaches\cite{Madkhali2025} demonstrate how surface functionalization can alter electronic properties, though adapting these to perovskite interfaces requires extensive investigation.
\par SCAPS-1D itself has inherent limitations as a one-dimensional simulator, unable to capture three-dimensional phenomena like grain boundary effects or non-uniform defect distributions. The dependency on input parameter quality is critical; parameters derived from Density Functional Theory (DFT) calculations vary with functional choice and methodological decisions. Growing concerns about parameter non-physicality in the literature- unrealistically low defect densities or excessive doping levels-allow simulations to breach the Shockley-Queisser limit, producing "record" efficiencies with no experimental basis. Translating simulated architectures to physical devices presents immense challenges. Synthesizing high-quality, phase-pure double perovskite films with controlled stoichiometry requires precise control over B-site cation ordering. Scalability of deposition methods remains a major hurdle, with industrial techniques requiring optimization for each composition. Economic viability and sustainability remain unproven, with some compositions requiring rare or expensive elements. The balance between high efficiency and low costs, fundamental to any photovoltaic technology, has yet to be convincingly demonstrated for lead-free double perovskites. As computational research on related materials grows from oxide perovskites for thermo-spintronics to ternary chalcogenides and MAX phases, lessons learned regarding synthesis challenges, stability issues, and the theory-experiment gap should guide development toward practical, commercially viable solutions.

\section{Strategies to overcome challenges and enhance efficiency} 
Compositional tuning of materials allows for optimizing the optoelectronic properties to the required state. For instance, monovalent doping (Li$^+$, Cu$^+$, Ag$^+$, etc.) can minimize trap states, leading to enhanced crystallinity and improved film quality, which enhance the PCE. Likewise, doping with bivalent cations (Zn$^{2+}$, Mn$^{2+}$, Co$^{2+}$) improves stability and trivalent cations (In$^{3+}$, Eu$^{3+}$, Al$^{3+}$) reduce defects while also refining the film morphology in addition to enhancing stability and PCE.  For better comprehension of the optoelectronic properties of perovskite, researchers often utilize theoretical techniques such as DFT. DFT can predict the electronic band structures, defect formation and their impact on the electronic properties, formation energy of defects, and carrier mobility. This allows researchers to minimize recombination loss due to trap states. Theoretical tools help experimental work in selecting suitable materials as well as fabricating a device.  
\par The stability issue of perovskite has been one of the main drawbacks in PSCs' commercialization. Investigating the degradation mechanisms using various characterization approaches such as spectroscopy, electron microscopy, and electrochemistry, in combination with theoretical studies, can help understand the deterioration of PSCs. Researchers incorporate methods including encapsulation, hydrophobic coatings, and replacing reactive metal electrodes with more stable materials like carbon and TCO to shield perovskite from environmental stressors. 
\par In order to obtain realistic simulation results using SCAPS-1D, researchers must acquire knowledge of SCAPS-1D and its inherent limitations. Varying parameters of the intrinsic properties of a material could lead to practically unattainable values. It is essential to use parameters that are experimentally feasible with realistic working conditions. Moreover, since SCAPS-1D does not account for optical losses and the default settings are oversimplified, it is crucial to incorporate scattering and interface phenomena along with reflection losses at the intermediate interface. Integration of multiple exciton generation, photon up-conversion, intermediate band photovoltaics, hot carrier cells and photon recycling in SCAPS-1D is also recommended.  
\section{Other Applications of Perovskites excluding solar cell} Beyond the scope of solar technology, double perovskite shows promising use in other optoelectronic applications.
\subsection{Photodetectors} In 2018, Luo \textit{et al.} successfully grew a high-quality Cs$_2$AgInCl$_6$ single crystal using a one-pot hydrothermal method, achieving an ultralow trap-state density (8.6 x 10$^8$ cm$^{-3}$). A stable Cs$_2$AgInCl$_6$-based UV photodetector with a high detective of 10$^{12}$ Jones, fast response speed of $\sim$ 1ms, high on/off ratio ($\sim 500$), and low dark current $\sim$ 10pA at 5V bias was fabricated.\cite{Lou2018} Lei \textit{et al.} prepared Cs$_2$AgBiBr$_6$ thin film by using one-step spin coating method. The device exhibit remarkable result: responsiveness of 7.01 AW$^{-1}$, on/off photocurrent ratio of 2.16 x 10$^4$, detective of 5.66 x 10$^{11}$ Jones, and fast photo-response of 956/995 $\mu$s. A notable discovery in this study is the superior stability demonstrated by the proposed Cs$_2$AgBiBr$_6$ against water and oxygen degradation over organic-inorganic hybrid perovskite Photodetectors when tested in ambient air.\cite{Lei2018}   
\subsection{Light-Emitting Diode} The journey of perovskite LEDs starts with a lead-based structure. In 2014, Tan \textit{et al.} introduced a solution-processed CH$_3$NH$_3$PbI$_{3-x}$Cl$_x$ emitter, demonstrating electroluminescence in the near-infrared, green, and red when tuning the halide compositions. The infrared device consists of TiO$_2$/CH$_3$NH$_3$PbI$_{3-x}$Cl$_x$/F8 sandwiched between indium tin oxide (ITO) and MoO$_3$/Ag, while the green device structure consists of PEDOT: PSS/TiO$_2$/CH$_3$NH$_3$PbI$_{3-x}$Cl$_x$/F8 between ITO and Ca/Ag. An infrared radiance of 13.2 W sr$^{-1}$ m$^{-2}$ at current density of 363 mA cm$^{-2}$ and a luminance of 364 cd m$^{-2}$ at 123 mA cm$^{-2}$ was achieved respectively.\cite{Tan2014} This offers a promising future for low-cost perovskite display, although toxic lead content restricts further commercial implementation. In 2018, Luo \textit{et al.} reported a lead-free double perovskite that emits stable white light. Alloying of Cs$_2$Ag$_{0.60}$Na$_{0.40}$InCl$_6$ plus 0.04\% bismuth doping emits a warm white light with QE of 86\% that works for 1,000 hours.\cite{Luo2018} Recently, a highly bright perovskite LED with peak radiance of 2409 W sr$^{-1}$ m$^{-2}$ and EQE of over 20\% at 2270 mA cm$^{-2}$ current density was reported. This notable finding is obtained by incorporating electron-withdrawing trifluoroacetate anions into a 3D perovskite emitter. This gives a reduction in Auger recombination due to a decoupled electron-hole wavefunction. These insights pave the way for utilizing perovskite emitters in high-power light-emitting technologies.\cite{Li2025}      
\subsection{Photocatalysis} The first successful water splitting using SrTiO$_3$ as a photocatalyst ignited the spark for perovskite photocatalysis.\cite{Wagner1980} In 2003, NaTaO$_3$ doped with La achieved high photocatalytic activity with an unprecedented apparent quantum yield (AQY) of 56\% at 270 nm.\cite{Kato2003} One of the most commonly studied perovskite materials in the context of photocatalysis is the CsPbBr$_3$ compound. To enhance its stability against humidity, Kong \textit{et al.} utilized ZIF coating on the surface of CsPbBr$_3$.\cite{Kong2018} Coating with a graphdiyne (GDY) layer also results in improved photocatalytic performance for CO$_2$ reduction.
In 2023, Fehr \textit{et al.} integrated a conductive adhesive barrier (CAB) in perovskite photoelectrochemical cells (PEC). Two different designs, namely, co-planar photocathode-photoanode architecture and silicon-perovskite tandem, exhibit STH efficiency of 13.4\% and 20.8\% respectively.\cite{Fehr2023} 
\subsection{Thermoelectric} Research on the thermoelectric (TE) properties of perovskite elevated after the first reported ultra-low thermal conductivity of MAPbI$_3$ in 2014.\cite{Pisoni2014, Haque2020} Due to their high mobility, diffusion length, and Seebeck coefficient in addition to low thermal conductivity, they are a promising candidate for TE applications. Recently, chalcogenide perovskite BaZrS$_3$ with a band gap of 1.9 eV was reported to have a ZT value of 0.81 at 750 K, the highest recorded among perovskites and other materials having a band gap higher than 1.5 eV. This was achieved by decoupling the electrical and thermal transport by alloying Se on S sites, which reduced the thermal conductivity while also maintaining the transport properties.\cite{Wu2026} By doping low-dimensional Cs$_3$Cu$_2$I$_5$ with barium, an increase in electrical conductivity by orders of 3 magnitude and a decrease in thermal conductivity was observed.\cite{Asker2025} One strategy among researchers in search of efficient TE materials is to utilize DFT and machine learning. Prediction of Rb$_2$AuCoF$_6$ and Rb$_2$AuRhF$_6$ TE performance using DFT and semiclassical Boltzmann transport theory revealed a ZT value of 0.177 and 0.13 with a Seebeck coefficient of 181$\mu$V/K, indicating comparatively promising use in TE applications.\cite{Yurong2025} 
\subsection{X-Ray Detectors} Stoumpos \textit{et al.} in 2013 laid the foundation for perovskite X-ray detectors with Cs$_2$PbBr$_3$ compound.\cite{Stoumpos2013} An advancement towards this field occurred when Wei \textit{et al.} revealed that MAPbBr$_3$ SC can be virtually integrated onto any substrate using solution-processed molecular bonding.\cite{Wei2017} In 2020, Tsai \textit{et al.} demonstrated MAPBI$_3$ thin film X-ray detector in p-i-n configuration, employing 2D Ruddlesden-Popper phase. This architecture significantly reduced the dark current and increased sensitivity with stable detection.\cite{Tsai2020} Jin \textit{et al.} recently developed Cs$_4$MnBi$_2$Cl$_{12}$:Tm/Ce scintillator for portable X-rays by using a hydrothermal process that provides clear internal imaging under bright conditions due to its efficient NIR emission. A light yield of 49964 photons MeV$^{-1}$ along with detection limit of 57.2 nGy$_{air}$s$^{-1}$ was reported. These results can be attributed to the 780 nm long-pass filter effectively blocking unwanted signals and the emission of the NIR-II window reducing background noise.\cite{Jin2026} To address the brittle nature of halide perovskite, Li \textit{et al.} adopted polymer-assisted crystalization and embedding method for Cs$_2$AgBiBr$_6$ (CABB) thick films for effective X-ray absorption. When modified with copolymer Pluronic P123, the sensitivity reached 244.71 $\mu$C Gy$^{-1}$ cm$^{-2}$ and low detection limit of 121 nGy s$^{-1}$. In addition, 70\% of the initial sensitivity was maintained for small bending cycles, and over 80\% at 500 bending cycles. A remarkable endurance and stability for more than 60 days was observed as compared to the prominent degradation observed in the parent CABB devices.           
\subsection{Piezoelectric} The piezoelectricity of perovskite materials was fist discovered in BaTiO$_3$ ceramics in 1946.\cite{Waqar2022} In 1954, Jaffe B. \textit{et al.} developed PZT ceramics that exhibits better piezoelectricity than BaTiO$_3$.\cite{Jaffe1954} Lead-free Bi$_{1/2}$Na$_{1/2}$TiO$_3$ ceramics emerged in 1960 discovered by Smolenski \textit{et al.}.\cite{ZHENG2018} In recent years, lead-free piezoelectric materials have received much attention driven by the global demand for toxic lead elimination. For instance, Venkatesan \textit{et al.} introduced ZnBr$_2$ in Cs$_3$Bi$_2$I$_9$ to stabilize it under ambient conditions and further confine the structure in a single crystal. The Cs$_3$Bi$_2$I$_9$:ZnBr$_2$ was incorporated into polyvinyl fluoride (PVDF) nanofiber, which results in better electroactive $\beta$ phase nucleation in PVDF, making it a promising material for piezoelectric nanogenerators.\cite{VENKATESAN2025} To enhance symmetry breaking in Cs$_2$AgBiBr$_6$, as its centrosymmetric structure (Fm$\bar{3}$m)causes poor piezoelectric response, Ong \textit{et al.} introduced 3-bromopropylammonium (BPA), n-butylammonium (BA), and phenylethylammonium (PEA) as barrier molecules to synthesise low-dimensional HDP. A d$_{33}$ value of 41.04 pm/V was achieved for quasi-2D PEA5, and the piezoelectric response increases from 2D to quasi-2D. An output linearity and sensitivity of 23.63 mV/kPa at 2.5-10 kPa was observed when fabricating a piezoelectric pressure sensor, comparable to PZT-based sensors, establishing their potential in piezoelectric applications.\cite{Ong2026} 

\section{Future Perspectives} The future outlook on perovskite centers on eliminating toxic lead as well as focusing on long-term stability for large-scale production and enhanced efficiency. It is necessary to develop recycling techniques to manage the environmental concerns for lead-based PSCs. To address the stability issue, an improved encapsulation that protects the device from any moisture is essential while also developing passivation methods to mitigate defects and finding new compounds through compositional techniques. Developing vapour-based deposition and optimization of solution-processing methods is pivotal for mass production. Furthermore, to decrease cost and guide experimental design, computational and simulation models must be augmented, as this will accelerate the production of stable and highly efficient PSCs. Further research and development should focus on addressing these future prospects to fully capture the potential of perovskite materials for the future of solar energy.

\section{Conclusions}
One of the most promising materials in the field of PV technology is PSCs, as they exhibit a PCE of over 26\%, comparable to the conventional Si-based solar cell. This paper provides an in-depth analysis of these materials regarding their fundamental properties, development, working mechanism, performance, challenges, and strategies to overcome limitations. Given the fact that lead-free double perovskites possess high optoelectronic potential, low toxicity, and stability, their practical efficiency has not yet reached the level of lead-based perovskite. We show that maximizing device performance, such as $>$32\% PCE for Cs$_2$CdPbI$_6$, demands a bandgap window of 1.5–1.8 eV, bulk defect densities well below 10$^{15}$ cm$^{-3}$, and high-mobility charge transport layers to limit non-radiative recombination. In this aspect, SCAPS--1D is very useful because it successfully simulates newer architectures, addressing the gap between theory and practice with efficiency, and it allows pinpointing the efficiency bottlenecks. But its inherent technical limitations are significant drawbacks. SCAPS-1D is a one-dimensional simulator and cannot simulate three-dimensional phenomena such as grain boundary effects or non-uniform defect distribution. Moreover, it is very sensitive to user-defined input parameters and can easily find efficiencies that are non-physical and higher than the theoretical limit when they are not carefully anchored to density functional theory data that is realistic.

\section*{Author contributions}
\begin{enumerate}
\item \textbf{H. Laltlanmawii:} Literature survey, Visualisation, Validation, Methodology, Final review writing \& editing. 
\item \textbf{Lalrem Kima:} Literature survey, Visualisation, Validation, Methodology, Formal analysis.
\item \textbf{Mahabur Rahman:} Literature survey, Visualisation, Validation, Methodology, Formal analysis. 
\item \textbf{Md. Ferdous Rahman:} Literature survey, Visualisation, Validation, Formal analysis. 
\item \textbf{Dilshod Nematov:} Literature survey, Visualisation, Validation, Methodology, Formal analysis. 
\item \textbf{S. Bhattarai:} Literature survey, Visualisation, Validation, Methodology, Formal analysis. 
\item \textbf{C.V.M. Chaturvedi:} Literature survey, Visualisation, Validation, Methodology, Formal analysis.
\item \textbf{Yazen M. Alawaideh:} Literature survey, Visualisation, Validation, Methodology, Formal analysis. 
\item \textbf{A. Laref:} Literature survey, Visualisation, Validation, Methodology.
\item \textbf{D. P. Rai:} Literature survey, Visualisation, Validation, Methodology \& Supervision. 
\end{enumerate}

\section*{Conflicts of interest}
There are no conflicts of interest. 

\section*{Data availability}

All data are included in the main text 

\section{Acknowledgments}
\textbf{A. Laref} acknowledges support from the "Research Center of the Female Scientific and Medical Colleges",  Deanship of Scientific Research, King Saud University.




\bibliography{rsc} 
\bibliographystyle{rsc} 
\end{document}